\documentclass[a4paper,11pt]{article}
\pdfoutput=1

\usepackage{jcappub}

\usepackage[T1]{fontenc}

\usepackage{amsmath}
\usepackage{amssymb}
\usepackage{amsfonts}
\usepackage{bm}
\usepackage{verbatim}
\usepackage{tablefootnote}
\usepackage{enumerate}
\usepackage{afterpage}
\usepackage{xcolor}
\usepackage{natbib, hyperref}

\usepackage{graphicx}
\usepackage{tabularx}
\usepackage{booktabs}
\usepackage{pifont}

\newcommand{\orcid}[1]{\href{https://orcid.org/#1}{\,\includegraphics[width=8px]{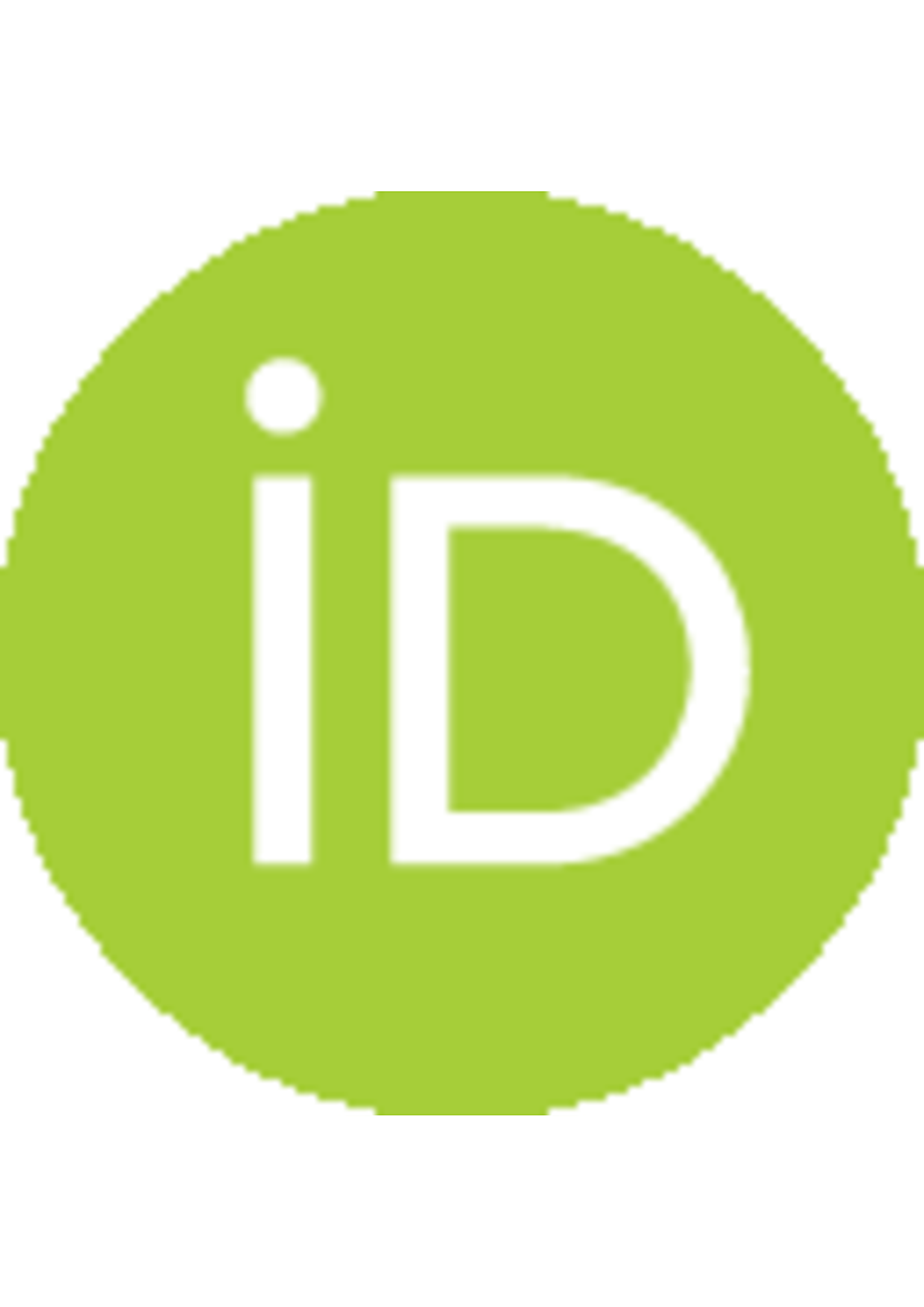}}}

\usepackage{soul}
\usepackage{xcolor}
\setstcolor{blue}

\usepackage{tikz}
\newcommand{\mathst}[1]{
  \tikz[baseline=(X.base)]{
    \node[inner sep=0pt, outer sep=0pt] (X) {$#1$};
    \draw[blue, line width=0.5pt] (X.west) -- (X.east);
}
}
\usepackage{etoolbox}
\robustify{\mathst}

\usepackage{makecell}

\title{\boldmath
Two powers, one scalar field: unified early and late dark energy
}

\author{
Bikash R. Dinda$^{a,b}$\orcid{0000-0001-5432-667X}
, Roy Maartens$^{a,b}$\orcid{0000-0001-9050-5894}
}

\affiliation[a]{Department of Physics \& Astronomy, University of the Western Cape, Cape Town 7535, South Africa}

\affiliation[b]{National Institute for Theoretical \& Computational Science, Cape Town 7535, South Africa}

\emailAdd{bikashrdinda@gmail.com}
\emailAdd{rmaartens@uwc.ac.za}

\abstract{
We investigate whether a single canonical scalar field can account for both early and late dark energy. We consider a double-power potential in which a steep positive-power term can generate an early peak in the scalar energy-density fraction, while an inverse-power term supports late-time acceleration. This evolution requires the steep term to dominate initially and become subdominant later. We test the model against cosmic microwave background (CMB) measurements, Dark Energy Spectroscopic Instrument (DESI) baryon acoustic oscillations, and Pantheon+ supernova data, with the SH0ES host-distance calibration. Relative to flat $\Lambda$CDM, the data fit improves by $\Delta\chi^2_{\rm data}=-6.90$, or $\Delta\chi^2_{\rm objective}=-3.76$ when Gaussian parameter-prior penalties are included. The scalar-field model gives $H_0=69.53\pm0.26\,{\rm km\,s^{-1}\,Mpc^{-1}}$, compared with $68.45\pm0.24\,{\rm km\,s^{-1}\,Mpc^{-1}}$ in flat $\Lambda$CDM. This shift brings the inferred expansion rate closer to the SH0ES measurement, but does not resolve the discrepancy significantly.
}
\notoc

\begin{document}
\maketitle
\flushbottom

\section{Introduction}
\label{sec:introduction}

The standard $\Lambda$CDM model in spatially flat Friedmann--Lemaître--Robertson--Walker (FLRW) spacetime successfully describes a wide range of cosmological observations. These include cosmic microwave background (CMB) temperature and polarization anisotropies \citep{Planck:2019nip,Planck:2018vyg}, baryon acoustic oscillation (BAO) measurements of cosmic distances \citep{BOSS:2016wmc,eBOSS:2020yzd,DESI:2025zgx}, luminosity distances measured with Type Ia supernovae (SNIa) \citep{SupernovaSearchTeam:1998fmf, SupernovaCosmologyProject:1998vns,Brout:2022vxf}, primordial deuterium and helium abundances predicted by Big Bang nucleosynthesis (BBN) \citep{Fields:2019pfx}, and measurements of structure growth from redshift-space distortions \citep{BOSS:2016wmc,eBOSS:2020yzd}. Despite these successes, the model leaves important theoretical questions unresolved. The vacuum-energy density inferred from observations is many orders of magnitude smaller than the scales suggested by particle physics. Explaining this small value is the cosmological-constant fine-tuning problem \citep{Weinberg:1988cp,Carroll:2000fy,Peebles:2002gy}. There is also no clear explanation for why the dark-energy and matter densities are comparable today, although their ratio changes 
as the Universe expands --  the cosmic-coincidence problem \citep{Carroll:2000fy,Peebles:2002gy,Velten:2014nra}.

In addition to these theoretical questions, $\Lambda$CDM faces observational tensions. A prominent example is the Hubble tension, which concerns the present expansion rate, $H_0$. Planck CMB observations, interpreted within spatially flat $\Lambda$CDM, give $H_0=67.4\pm0.5\,\mathrm{km\,s^{-1}\,Mpc^{-1}}$ \citep{Planck:2018vyg}. In contrast, the SH0ES Cepheid-calibrated SNIa distance ladder gives $H_0=73.04\pm1.04\,\mathrm{km\,s^{-1}\,Mpc^{-1}}$ \citep{Riess:2021jrx}. The quoted uncertainties are at $68\%$ confidence, and these determinations differ by approximately $5\sigma$. Riess et al.\ compared stellar distances measured with the James Webb Space Telescope (JWST) and the Hubble Space Telescope (HST). They found that the HST Cepheid distances agree with the other distance estimates at approximately $1\sigma$ \citep{Riess:2024vfa}. This comparison provides no evidence for a distance bias large enough to explain the Hubble tension. However, different distance-ladder calibrations give different values of $H_0$. Freedman used the tip of the red giant branch (TRGB) to calibrate SNIa and obtained $H_0=69.8\pm0.6\,(\mathrm{stat})\pm1.6\,(\mathrm{sys})\, \mathrm{km\,s^{-1}\,Mpc^{-1}}$ \citep{Freedman:2021ahq}. Combining these uncertainties in quadrature, this value differs from the Planck $\Lambda$CDM inference by approximately $1.3\sigma$. The significance of the Hubble tension therefore depends on the local calibration used. Its origin remains unresolved \citep{DiValentino:2021izs,Schoneberg:2021qvd, Abdalla:2022yfr,Dong:2026epd}. For other aspects related to the Hubble tension, see \citep{Ling:2026awm,Najera:2026odk}.

Another observational tension concerns the amplitude of matter clustering. It is commonly expressed through
\begin{equation}
    S_8\equiv \sigma_8\left(\frac{\Omega_{\rm m0}}{0.3}\right)^{1/2},
\label{eq:S8_definition}
\end{equation}
where $\sigma_8$ is the present root-mean-square linear matter density contrast smoothed with a spherical top-hat filter of radius $8\,h^{-1}\,\mathrm{Mpc}$. Here, $\Omega_{\rm m0}$ is the present matter density parameter, and $h=H_0/(100\,\mathrm{km\,s^{-1}\,Mpc^{-1}})$. Several weak-lensing and galaxy-clustering analyses have inferred lower values of $S_8$ than CMB observations within $\Lambda$CDM \citep{Heymans:2020gsg,DES:2021wwk,Li:2023tui}. This discrepancy is commonly called the $S_8$ tension \citep{DiValentino:2020vvd,Abdalla:2022yfr,Pantos:2026koc}. Recent results show that its significance depends on the datasets and the statistical comparison. The Dark Energy Survey (DES) Year~6 joint analysis of cosmic shear, galaxy clustering, and galaxy--galaxy lensing gives $S_8=0.789\pm0.012$ in spatially flat $\Lambda$CDM \citep{DES:2026fyc}. The same paper independently analyses primary CMB data from Planck~2018, Atacama Cosmology Telescope Data Release~6 (ACT DR6), and South Pole Telescope third-generation camera Data Release~1 (SPT-3G DR1). This CMB-only combination gives $S_8=0.836^{+0.012}_{-0.013}$. Comparing the one-dimensional marginalised $S_8$ posteriors, the DES collaboration reports a $2.6\sigma$ discrepancy between DES Year~6 and this CMB combination \citep{DES:2026fyc}. In contrast, the complete Kilo-Degree Survey (KiDS Legacy) cosmic-shear analysis gives $S_8=0.815^{+0.016}_{-0.021}$ and reports a $0.73\sigma$ difference from Planck alone \citep{Wright:2025xka}. Thus, current weak-lensing results do not show a uniform preference for less matter clustering than inferred from the CMB \citep{DES:2026fyc,Wright:2025xka,Pantos:2026koc}. For other related recent studies see \citep{Jones:2026ndt}.

The CMB inference of $H_0$ depends on the assumed cosmological model. One of the most precisely constrained quantities is the angular acoustic scale
\citep{Planck:2018vyg},
\begin{equation}
\theta_*=\frac{r_s(z_*)}{D_M(z_*)},
\label{eq:angular_acoustic_scale}
\end{equation}
where $z_*$ is the redshift of photon last scattering. The comoving sound horizon at this epoch is
\begin{equation}
    r_s(z_*)=
    \int_{z_*}^{\infty}\frac{c_s(z)}{H(z)}\,dz,
\label{eq:sound_horizon}
\end{equation}
where $c_s(z)$ is the sound speed of the photon--baryon fluid \citep{Hu:1995en,Planck:2018vyg}. In a spatially flat universe, the transverse comoving distance to last scattering is
\citep{Hogg:1999ad}
\begin{equation}
    D_M(z_*)=
    \frac{c}{H_0}\int_0^{z_*}\frac{dz}{E(z)},
    \qquad
    E(z)\equiv\frac{H(z)}{H_0}.
\label{eq:distance_last_scattering}
\end{equation}
\autoref{eq:distance_last_scattering} shows that one possible way to accommodate a higher $H_0$ is through a smaller distance $D_M(z_*)$. Preserving the observed angular acoustic scale in \autoref{eq:angular_acoustic_scale} then requires a smaller sound horizon $r_s(z_*)$. One way to achieve this is by increasing the expansion rate before last scattering, as indicated by \autoref{eq:sound_horizon} \citep{Bernal:2016gxb,Aylor:2018drw,Knox:2019rjx}. This motivates early dark energy (EDE), an additional component that briefly contributes a non-negligible fraction of the total energy density before recombination and then dilutes rapidly \citep{Poulin:2018cxd,Smith:2019ihp,Poulin:2023lkg}. For other recent studies on EDE see \citep{Giare:2026gdn}.

A canonical scalar field can produce an EDE epoch. At early times, Hubble friction can keep the field nearly frozen, so its energy density remains approximately constant. Its fraction of the total density then increases as radiation and matter dilute. As the expansion rate decreases, the field can roll down its potential, and its equation of state can increase from $w_\phi\simeq-1$. The scalar fraction reaches a peak if its density subsequently begins to dilute faster than the combined density of the other components \citep{Poulin:2018cxd,Smith:2019ihp}. Such behaviour has been studied with axion-like and power-law potentials \citep{Poulin:2018cxd,Agrawal:2019lmo,Smith:2019ihp}. Related proposals include acoustic dark energy \citep{Lin:2019qug}, first-order phase transitions \citep{Niedermann:2019olb,Niedermann:2020dwg}, and scalar fields whose evolution is influenced by couplings to matter \citep{Karwal:2021vpk,McDonough:2021pdg,Lin:2022phm}.

The viability of this mechanism depends on more than the reduction in the sound horizon. EDE also changes gravitational potentials, photon diffusion, and the evolution of cosmological perturbations. These changes affect the heights and shapes of the CMB acoustic peaks and the damping tail \citep{Smith:2019ihp,Knox:2019rjx,Vagnozzi:2021gjh, Poulin:2023lkg}. Preserving $\theta_*$ alone therefore does not preserve the observed CMB spectra. EDE models must be tested against CMB temperature, polarization, and lensing observations \citep{Poulin:2018cxd,Smith:2019ihp,Smith:2023oop, Efstathiou:2023fbn}.

CMB and galaxy-clustering analyses have placed strong limits on the allowed early contribution \citep{Hill:2020osr,Ivanov:2020ril,DAmico:2020ods, Efstathiou:2023fbn,Gsponer:2023wpm}. Weak-lensing and Lyman-$\alpha$ forest measurements provide further tests \citep{Murgia:2020ryi,Goldstein:2023gnw}. However, marginalized posterior distributions and profile likelihoods can give different conclusions about the preferred EDE amplitude \citep{Herold:2021ksg,Herold:2022iib,Poulin:2025nfb}. Recent analyses using ACT DR6 and DESI DR2, and combinations including galaxy-cluster abundances, further illustrate the dependence on the adopted observations \citep{Poulin:2025nfb,Strunk:2026vve}. The broader status of EDE models and their observational constraints is reviewed in \citep{Poulin:2023lkg,Khalife:2023qbu,Giare:2024akf}. Moreover, a higher inferred $H_0$ does not necessarily come with a lower $S_8$. In some EDE models, the changes in matter density and primordial perturbations required to fit the CMB increase the predicted clustering amplitude \citep{Hill:2020osr,Ivanov:2020ril,Murgia:2020ryi}. The implications for the Hubble and $S_8$ tensions must therefore be assessed separately.

In standard EDE constructions, the early scalar contribution becomes negligible after its peak, and a separate cosmological constant drives late-time acceleration \citep{Poulin:2018cxd,Smith:2019ihp}. Other studies introduce separate dynamical components for early and late dark energy \citep{Jhaveri:2026bla}. An alternative is to ask whether a single scalar field can produce both an early epoch and the observed late-time acceleration. In this case, the field must first develop a peak in its energy-density fraction. Its fraction must then decrease sufficiently to preserve the subsequent matter-dominated era. At late times, the same field must contribute enough energy with sufficiently negative pressure to drive acceleration \citep{Brissenden:2023yko,Sohail:2024oki,Giare:2026tyk}.

Several studies have explored this possibility. Examples include $\alpha$-attractor potentials with a constant offset for late acceleration \citep{Braglia:2020bym}, non-oscillating EDE followed by late-time quintessence \citep{Brissenden:2023yko}, a modified steep-exponential potential \citep{Sohail:2024oki}, and a localized potential feature with an exponential tail \citep{Kouniatalis:2026upo}. Other approaches use successive vacuum transitions \citep{Freese:2021rjq}, matter-coupled scalar dynamics \citep{Ramadan:2023ivw}, or matter-dependent transitions between early and late dark-energy regimes \citep{Guendelman:2025swp}. Broader constructions also connect inflation with early and late dark energy through multiple scalar fields or gravitational couplings \citep{Guendelman:2022cop,Nojiri:2023mvi}. These studies motivate testing whether a double-power potential can produce an early episode and late-time acceleration along a single scalar-field trajectory.

We investigate this question with a double-power potential for a single canonical scalar field. The potential contains a steep positive-power term and an inverse-power term. The steep term is intended to control the early evolution, while the inverse-power term becomes dominant later and supports late-time acceleration. Both epochs follow from the same potential and the same scalar-field trajectory. The ordering of the powers and the initial dominance of the steep term do not by themselves guarantee the required cosmological history. An early peak requires the scalar energy-density fraction to first increase and then decrease. Late-time acceleration requires a separate check of the field's density and pressure. We examine these conditions using the background evolution and then include the scalar perturbations in the Einstein--Boltzmann calculation.

We test the model against CMB temperature, polarization, and lensing measurements, DESI Data Release~2 BAO, and Pantheon+ supernova observations with the SH0ES host-distance calibration. We compare the constraints and fit quality with those of spatially flat $\Lambda$CDM. With the potential powers and initial field conditions fixed, the model introduces only one additional free parameter relative to flat $\Lambda$CDM.

We use the metric signature $(-,+,+,+)$ and units with $c=\hbar=1$, where $c$ is the speed of light and $\hbar$ is the reduced Planck constant.
The reduced Planck mass is $M_{\rm Pl}=(8\pi G)^{-1/2}$, where $G$ is Newton's gravitational constant.

\section{Brief overview of quintessence dynamics}
\label{sec:quintessence_dynamics}

We consider a canonical scalar field $\phi$ minimally coupled to gravity, with potential $V$. The field interacts with the other cosmological components only through gravity. The action of the complete system is $S=S_\phi+S_{\rm other}[g_{\mu\nu},\Psi_i]$, where
\begin{equation}
    S_\phi
    =\int d^4x\,\sqrt{-g}
    \left[-\frac12 g^{\mu\nu}\nabla_\mu\phi\nabla_\nu\phi
        -V(\phi)\right] \,.
\label{eq:scalar_action}
\end{equation}
Varying \autoref{eq:scalar_action} with respect to 
$g^{\mu\nu}$ gives the scalar energy--momentum tensor,
\begin{equation}
    T_{\mu\nu}^{(\phi)}
    \equiv
    -\frac{2}{\sqrt{-g}}
    \frac{\delta S_\phi}{\delta g^{\mu\nu}}
    =    \nabla_\mu\phi\nabla_\nu\phi
    -g_{\mu\nu}
    \left[        \frac12\nabla_\lambda\phi\nabla^\lambda\phi
        +V(\phi)
    \right] \,.
\label{eq:scalar_energy_momentum}
\end{equation}
Varying the action
with respect to $\phi$ gives the Klein--Gordon equation,
\begin{equation}
    \Box\phi-V_{,\phi}=0,
    \qquad
    \Box\equiv\nabla_\mu\nabla^\mu.
\label{eq:covariant_kg}
\end{equation}
This ensures separate conservation of the scalar
energy--momentum tensor,
$\nabla_\mu T^{\mu\nu}_{(\phi)}=0$.


In a spatially flat FLRW
background, 
    $ds^2=a^2(\tau)
    \left(-d\tau^2+ d\bm{x}^2
    \right)$,
where $\tau$ is conformal time and $a$ is the scale factor. A prime denotes $d/d\tau$, and a dot denotes $d/dt$, where $dt=a\,d\tau$. The conformal Hubble rate is $\mathcal{H}\equiv {a'}/{a}=aH=\dot a$.
For the homogeneous field $\bar{\phi}(\tau)$, \autoref{eq:scalar_energy_momentum} gives energy density and pressure as
\begin{equation}    \rho_\phi=\frac{\bar{\phi}'^2}{2a^2}+V(\bar{\phi}) \,,
    \qquad
    p_\phi=\frac{\bar{\phi}'^2}{2a^2}-V(\bar{\phi}) \,,
\label{eq:scalar_density_pressure}
\end{equation}
with equation of state 
    $w_\phi={p_\phi}/{\rho_\phi}$.
The background Klein--Gordon equation is 
\begin{equation}
    \bar{\phi}''
    +2\mathcal{H}\bar{\phi}'
    +a^2 V_{,\bar{\phi}}=0 \,.
\label{eq:background_kg_conformal}
\end{equation}


Linear scalar perturbations in synchronous gauge 
lead to \citep{Ma:1995ey} 
\begin{equation}
    ds^2=a^2(\tau)
    \left[
        -d\tau^2+
        \left(\delta_{ij}+\gamma_{ij}\right)dx^i dx^j
    \right],
\label{eq:synchronous_metric}
\end{equation}
where 
\begin{equation}
    \gamma_{ij}(\bm{x},\tau)
    =
    \int\frac{d^3k}{(2\pi)^3} \, {\rm e}^{{\rm i}\,\bm{k}\cdot\bm{x}}
    \left[\hat{k}_i\hat{k}_j\, h(\bm{k},\tau)
        +6\eta(\bm{k},\tau)
        \left(
            \hat{k}_i\hat{k}_j-\frac13\delta_{ij}
        \right)
    \right],
\label{eq:scalar_metric_decomposition}
\end{equation}
with $\hat{k}_i=k_i/k$ and $k=|\bm{k}|$. 
Then \autoref{eq:covariant_kg} gives
\begin{equation}
    \delta\phi''
    +2\mathcal{H}\delta\phi'
    +\left(k^2+a^2V_{,\bar{\phi}\bar{\phi}}\right)\delta\phi
    =
    -\frac12\bar{\phi}'h'\,,
\label{eq:perturbed_kg}
\end{equation}
where the right-hand side describes the coupling to the scalar metric perturbations. 
Linearising \autoref{eq:scalar_energy_momentum} gives the scalar energy density and pressure perturbations,
\begin{equation}
    \delta\rho_\phi
    =    \frac{\bar{\phi}'\delta\phi'}{a^2}
    +V_{,\bar{\phi}}\delta\phi,
    \qquad
    \delta p_\phi
    =    \frac{\bar{\phi}'\delta\phi'}{a^2}
    -V_{,\bar{\phi}}\delta\phi\,,
\label{eq:scalar_density_pressure_perturbations}
\end{equation}
the 
momentum perturbation,
\begin{equation}
    \delta T^0_{(\phi)\,i}
    =
    -{\rm i}\,\frac{ k_i}{a^2}\bar{\phi}'\delta\phi \,,
    \label{eq:scalar_momentum}
\end{equation}
and the velocity divergence,
\begin{equation}
    (\bar{\rho}_\phi+\bar{p}_\phi)\theta_\phi
    \equiv {\rm i}\, k^i\delta T^0_{(\phi)\,i}
    =
    \frac{k^2}{a^2}\bar{\phi}'\delta\phi \, .
\label{eq:scalar_velocity_divergence}
\end{equation}
%
A canonical scalar has no linear anisotropic stress.
The sound speed is defined in the physical rest frame, which is given by $\delta\phi=0$. By \autoref{eq:scalar_density_pressure_perturbations} this leads to
\begin{equation}
    c_{s,\phi}^2
    \equiv
    \left.
    \frac{\delta p_\phi}{\delta\rho_\phi}
    \right|_{\rm rest}
    =1 \,.
\label{eq:scalar_sound_speed}
\end{equation}

We solve the coupled Einstein and conservation equations using a modified version of the Cosmic Linear Anisotropy Solving System (\texttt{CLASS}), accessed through its Python interface, \texttt{classy} \citep{Lesgourgues:2011re,Blas:2011rf}. 
%
We calculate the linear matter power spectrum with \texttt{CLASS}, including the scalar-field perturbations. To account for nonlinear matter clustering, we use a modified implementation of \texttt{HMcode-2020} \citep{Mead:2020vgs}\footnote{The modifications accommodate the scalar-field background history in the nonlinear prescription, which also uses the linear matter power spectrum computed by \texttt{CLASS}. We have not validated this extension against simulations for the double-power model. Such a validation lies beyond the scope of this work. The resulting parameter constraints and fit comparisons may depend on the nonlinear modelling. The background mechanism for an early scalar contribution followed by late acceleration does not rely on this prescription.}. 
We access these calculations through the Python interface \texttt{classy}.

\begin{figure}[!htbp]
\centering
\includegraphics[width=0.49\linewidth]{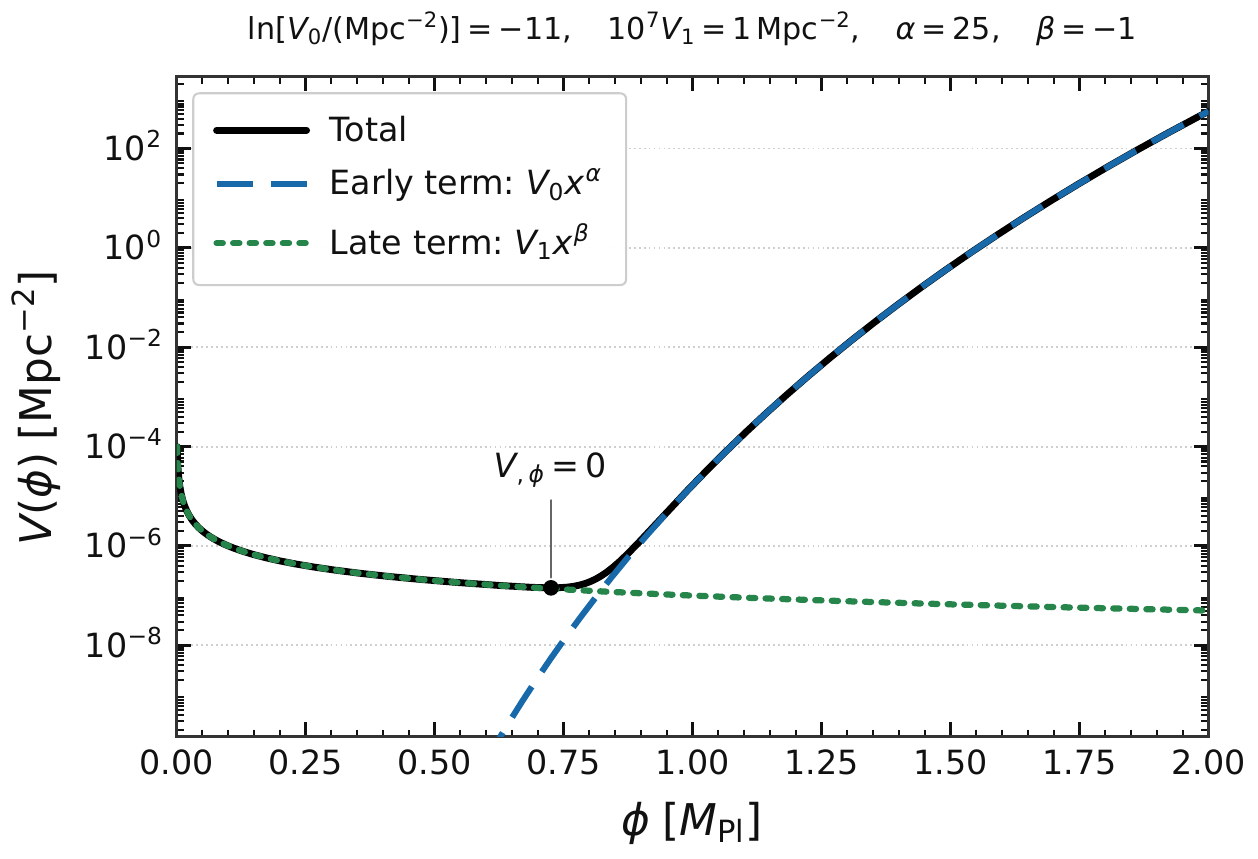}
\includegraphics[width=0.49\linewidth]{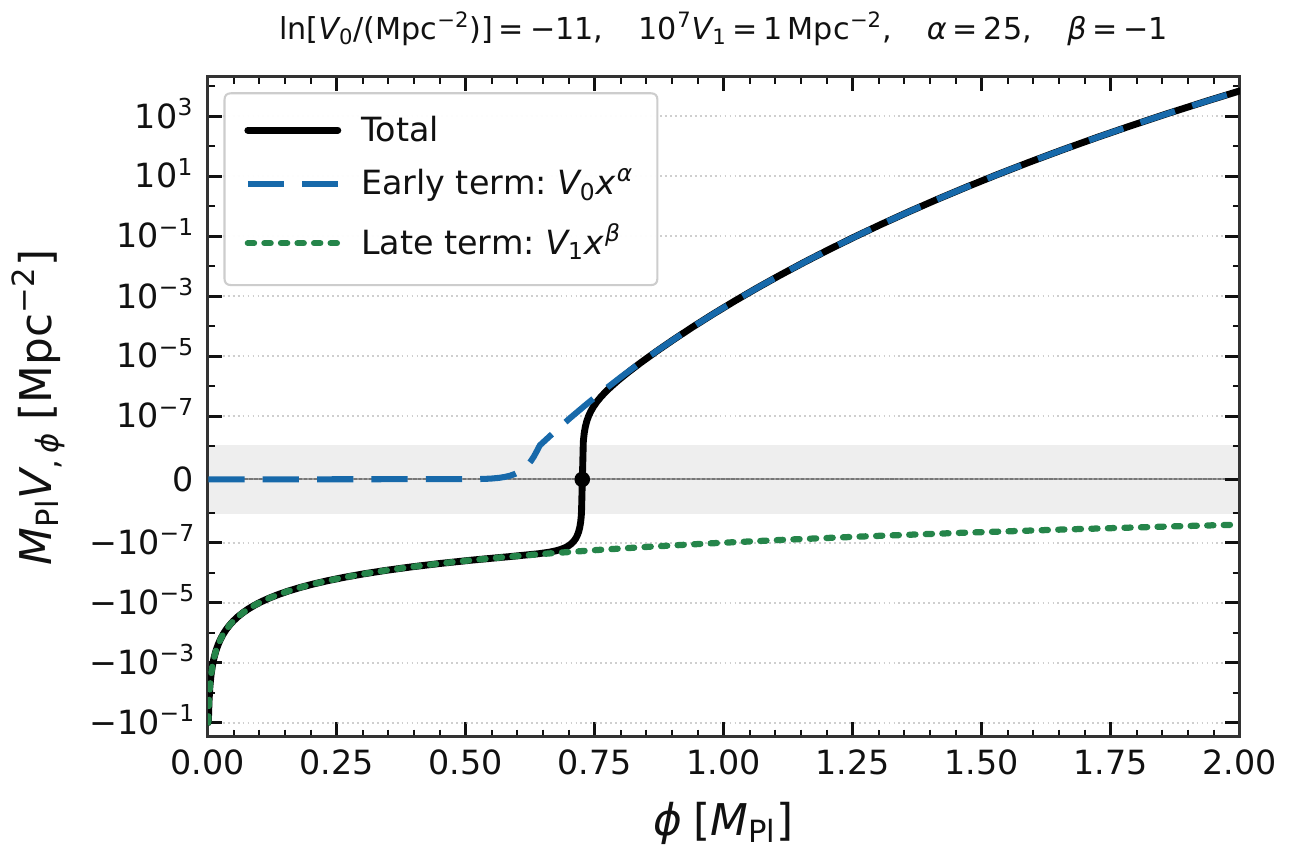} \\
\includegraphics[width=0.49\linewidth]{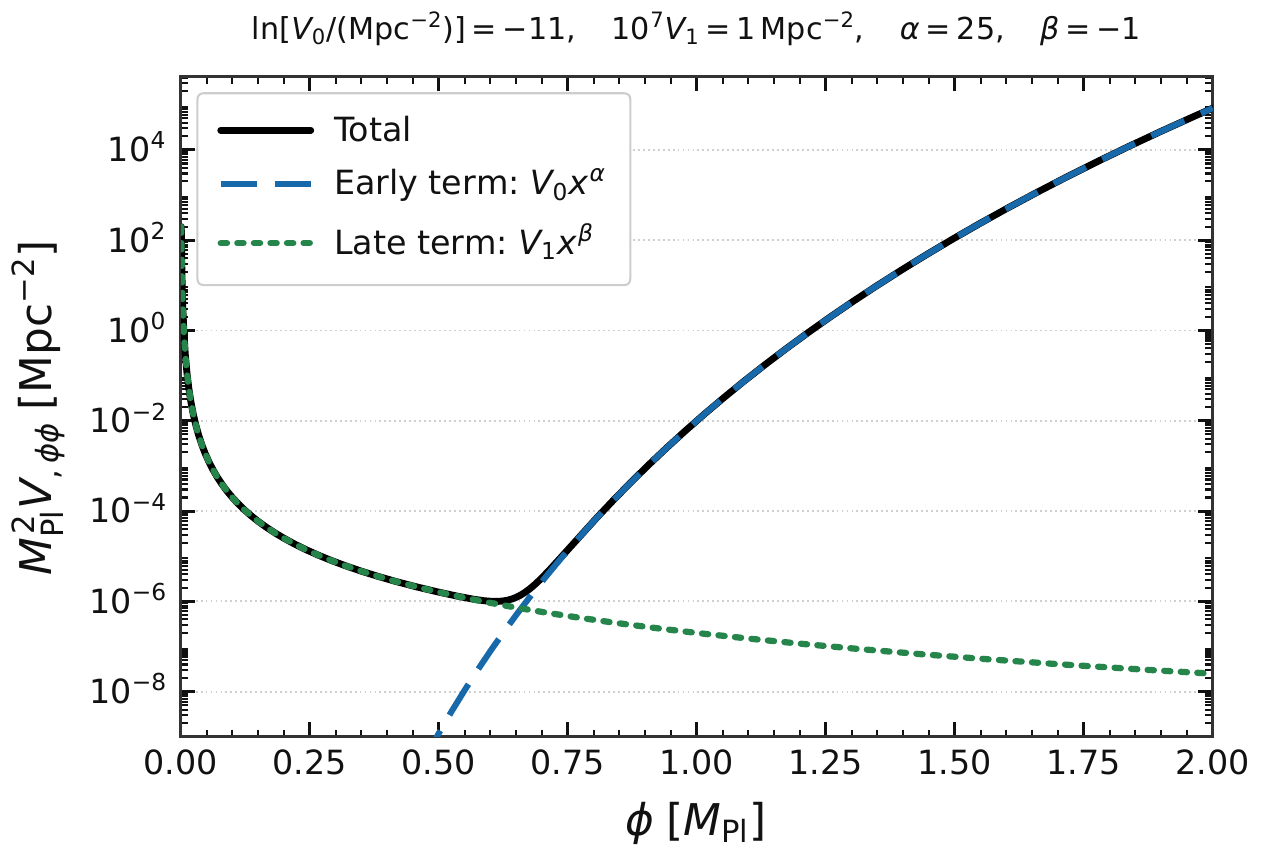}
\caption{The double-power potential (top-left), its first derivative (top-right), and its second derivative (bottom). The potential  is expressed in  \texttt{CLASS}'s internal units. The first derivative uses a signed logarithmic scale. Curves are evaluated for $10^{-3}\leq x\leq2$, since the inverse-power term is singular at $x=0$.
}
\label{fig:potential_derivatives}
\end{figure}

\section{The double-power potential}
\label{sec:potential}

We introduce the potential
\begin{equation}
    V(\phi)
    =
    V_0\left(\frac{\phi}{M_{\rm Pl}}\right)^\alpha
    +
    V_1\left(\frac{\phi}{M_{\rm Pl}}\right)^\beta,
    \qquad V_0,V_1>0,
\label{eq:double_power}
\end{equation}
where $\alpha$ and $\beta$ are dimensionless powers. The coefficients $V_0$ and $V_1$ have dimensions of energy
density. We restrict the field to $\phi>0$, so that $V$
is well defined.

In \autoref{eq:double_power}, the $V_0$ term is intended to dominate at early times, whereas the $V_1$ should dominate at low redshifts and drive the late-time cosmic acceleration. For a field rolling towards smaller positive values, we choose
\begin{equation}
    \alpha>0,
    \qquad
    \alpha>\beta \,.
    \label{eq:power_ordering}
\end{equation}
This ordering does not yet guarantee an early peak followed by a late-time acceleration, which we discuss later.
A single-power plus a constant term potential can produce a transient early contribution, as illustrated by the `rock-and-roll' scenario \citep{Agrawal:2019lmo}. In \autoref{eq:double_power}, the additional power provides a possible late-time deviation from the cosmological constant. No second scalar field is introduced. The early and late era therefore share one background trajectory and one set of scalar-field perturbations.
%
The potential derivatives 
are
\begin{align}
    V_{,\phi}
   & =
    \frac{1}{M_{\rm Pl}}
    \left[
     \alpha V_0x^{\alpha-1}
     +\beta V_1x^{\beta-1}
    \right]\quad \text{with}\quad x\equiv\frac{\phi}{M_{\rm Pl}}\,,   \label{eq:potential_first_derivative}
\\
    V_{,\phi\phi}
    &=
    \frac{1}{M_{\rm Pl}^2}
    \left[    \alpha(\alpha-1)V_0x^{\alpha-2}
    +\beta(\beta-1)V_1x^{\beta-2}
    \right].   \label{eq:potential_second_derivative}
\end{align}

\autoref{fig:potential_derivatives} shows the potential and its derivatives for an illustrative parameter choice. The steep $V_0x^{25}$ term dominates at larger field values, whereas the inverse-power term $V_1x^{-1}$ dominates at smaller values. Their competing slopes produce a minimum at $\phi\simeq0.7M_{\rm Pl}$, where $V_{,\phi}=0$. These plots describe the potential shape; whether the field produces an early peak and late acceleration depends on its cosmological evolution, which we shall see later. Note that the second derivative is always positive, and this will be useful to keep in mind for our later discussion.

\subsection{Initial conditions and potential-term dominance}
\label{subsec:potential_dominance}

We adopt the frozen initial condition,
\begin{equation}
    \bar{\phi}'_{\rm ini}=0 \,, \label{eq:scalar_initial_conditions}
\end{equation}
so that $w_{\phi,\rm ini}=-1$. We choose
    $x_{\rm ini} = 2$ 
for simplicity, although we could vary it as a sampled parameter. 
In order to compare the two contributions in
\autoref{eq:double_power}, we define
\begin{equation}
    \mathcal{R}(x)
    \equiv
    \frac{V_0x^\alpha}{V_1x^\beta}
    =
    \frac{V_0}{V_1}x^{\alpha-\beta}\,,
\label{eq:potential_ratio}
\end{equation}
so that the $V_0$ term dominates when $\mathcal{R}>1$, and the $V_1$ term dominates when
$\mathcal{R}<1$.

For the early evolution to be controlled by $V_0x^\alpha$ in \autoref{eq:double_power}, this term and its derivative must both dominate at early time compared to the other term: 
\begin{equation}
    \mathcal{R}_{\rm ini}\gg1,
    \qquad
    |\alpha|\mathcal{R}_{\rm ini}\gg|\beta| \,.
\label{eq:initial_energy_force_dominance}
\end{equation}
Since $\alpha>0$, dominance of the $V_0$ contribution 
requires $V_{,\phi}(\phi_{\rm ini})>0$. The potential therefore drives the field towards smaller $\phi$. At sufficiently early times, however, the large expansion rate suppresses its motion. The field remains approximately frozen, with nearly constant $\phi$ and negligible kinetic energy. As $H$ decreases, the field can move appreciably down the potential, and its kinetic energy can grow relative to its potential energy. The exact condition for this initial direction is
\begin{equation}
    V_{,\phi}(\phi_{\rm ini})>0 \quad\Longleftrightarrow\quad
    \alpha\mathcal{R}_{\rm ini}+\beta>0 \,.
\label{eq:initial_roll_direction}
\end{equation}

As the field decreases, the relative contribution of $V_0x^\alpha$ changes according to
\begin{equation}
    \frac{d\ln\mathcal{R}}{d\ln a}
    =
    (\alpha-\beta)\frac{d\ln x}{d\ln a}.
    \label{eq:ratio_evolution}
\end{equation}
For $\alpha>\beta$, decreasing $x$ therefore reduces $\mathcal{R}$.
The two  contributions are equal at
\begin{equation}
    x_\times
    =
    \left(\frac{V_1}{V_0}\right)^{1/(\alpha-\beta)}.
\label{eq:potential_crossover}
\end{equation}
The $V_0$ term dominates for $x>x_\times$, and the $V_1$  for $x<x_\times$. Thus a transition from early $V_0$ dominance to late $V_1$ dominance requires the field to cross $x_\times$ from above.

If $x_L$ denotes the field value at a specified late epoch, dominance of the intended term at each epoch
requires
    $x_{\rm ini}>x_\times>x_L$.
For $x_{\rm ini}=2$, the initial ratio and crossover
condition become
\begin{equation}
    \mathcal{R}_{\rm ini}
    =
    \frac{V_0}{V_1}2^{\alpha-\beta},
    \qquad
    2>x_\times>x_L.
\label{eq:fixed_initial_special_case}
\end{equation}

The initial inequalities in
\autoref{eq:initial_energy_force_dominance} allow the early motion to be approximated using $V_0x^\alpha$. The subsequent takeover by $V_1x^\beta$ requires the evolved trajectory to reach $x<x_\times$. Neither condition alone guarantees an early peak in the energy density parameter  $\Omega_\phi$. For the scalar field to have an early peak still requires  extra conditions, which we discuss later.

\begin{figure}[!htbp]
\centering
\includegraphics[width=0.9\linewidth]{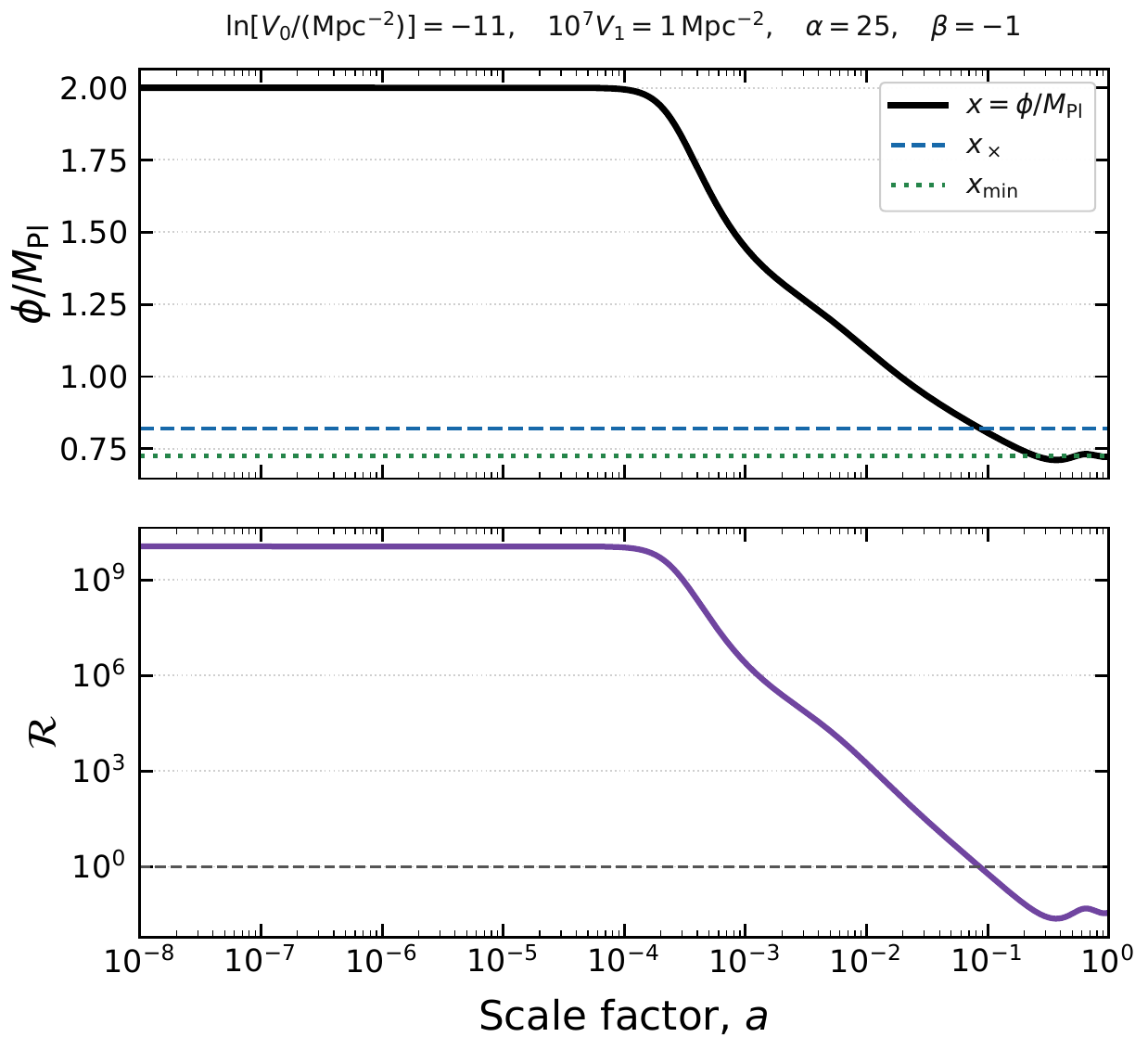}
\caption{
Evolution of the field (upper panel) and the ratio of the two potential contributions (lower panel). The horizontal lines in the upper panel mark equality of the potential contributions, $x_\times\simeq0.8$, and the potential minimum, $x_{\min}\simeq0.7$. In the lower panel, $\mathcal R>1$ denotes dominance of $V_0x^\alpha$, and $\mathcal R<1$ denotes dominance of $V_1x^\beta$. We use $x_{\rm ini}=2$, zero initial field velocity, $\omega_b=0.0225$, $\omega_{\rm cdm}=0.12$, and other standard early contributions.
}
\label{fig:field_dominance}
\end{figure}

\subsection{Late-time acceleration and thawing}
\label{subsec:late_acceleration}

When the $V_1$ term in \autoref{eq:double_power} becomes dominant, it can support late-time acceleration\footnote{$V_1$ domination is not necessary  for late-time cosmic acceleration. $V_0$ domination can also contribute to late-time acceleration. For simplicity, we retain the late-time dominance by $V_1$.}. 
The field must have sufficiently negative pressure and contribute enough to the total energy density to satisfy $\ddot{a}>0$. Potential-term dominance by itself does not establish these conditions.

For $\beta<0<\alpha$, the full potential has a positive minimum at $V_{,\phi}=0$:
\begin{equation}
    x_{\min}
    =
    \left(
    \frac{-\beta V_1}{\alpha V_0}
    \right)^{1/(\alpha-\beta)} 
    \quad \text{with} \quad
    V_{,\phi\phi}(\phi_{\min})
    =
    \frac{\alpha(\alpha-\beta)V_0}{M_{\rm Pl}^2}\,
    x_{\min}^{\alpha-2}>0 
    \,.
\label{eq:minimum_curvature}
\end{equation}
If the field settles at this minimum, its density approaches a positive constant and $w_\phi\rightarrow-1$. This is a property of the combined potential, rather than of an isolated inverse-power term.
A late thawing phase requires the field to remain nearly frozen before moving away from $w_\phi\simeq-1$ \citep{Caldwell:2005tm}. For a trajectory with a significant early energy density contribution, this requires an intermediate period in which the kinetic energy becomes negligible relative to the potential. If the field subsequently resumes rolling, the ratio of kinetic energy/potential increases and $w_\phi$ rises above $-1$.

Thus, depending on the potential parameters and the field's position and velocity after the early era, the field can remain nearly frozen until today or resume rolling at late times. If it eventually settles at the potential minimum, its equation of state approaches $w_\phi=-1$.
\autoref{fig:field_dominance} shows that the field initially changes little. It subsequently rolls towards smaller values and crosses $x_\times$. The inverse-power contribution then exceeds the positive-power contribution. The field then moves around the minimum of the combined potential.

\begin{figure}[!htbp]
\centering
\includegraphics[width=0.9\linewidth]{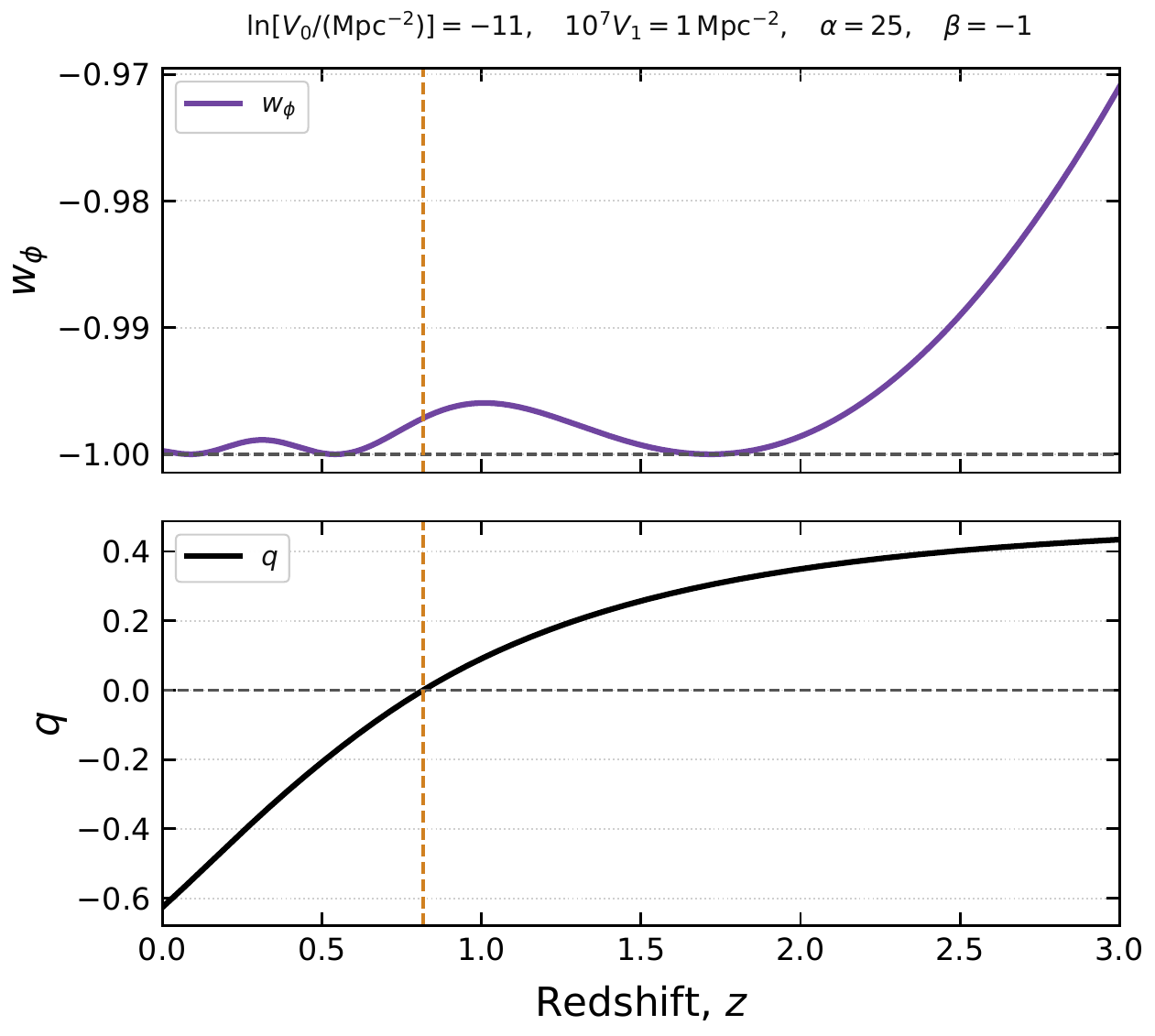}
\caption{Late evolution of the scalar equation of state (upper panel) and  deceleration parameter $q=-\ddot a/(aH^2)$ (lower panel). The horizontal dashed lines mark $w_\phi=-1$ and $q=0$. The vertical line marks the onset of acceleration at $z\simeq0.8$.}
\label{fig:late_acceleration}
\end{figure}

For this solution, the Universe begins to accelerate at $z\simeq0.8$, when $q$ changes from positive to negative, as shown in \autoref{fig:late_acceleration}.
At low redshifts, $w_\phi$ stays close to $-1$ but shows
small oscillations, repeatedly reaching $-1$.

\section{Conditions for an early peak in the scalar-field
energy-density fraction}
\label{sec:early_peak}

An early dark energy peak is a local maximum of the scalar energy-density fraction, $\Omega_\phi$. It is not a maximum of the scalar density $\rho_\phi$, which cannot increase in an expanding universe. We first derive the condition for this maximum and then discuss how it can arise in \autoref{eq:double_power}.


For the spatially flat universe considered here,
\begin{equation}
    \Omega_\phi
    =
    \frac{\rho_\phi}{\rho_\phi+\rho_{\rm other}},
    \qquad
    w_{\rm other}
    =
    \frac{p_{\rm other}}{\rho_{\rm other}},
\label{eq:scalar_fraction}
\end{equation}
where $\rho_{\rm other}$ and $p_{\rm other}$ are the combined density and pressure of all non-scalar components. Writing $N=\ln a$, separate energy conservation gives
\begin{equation}
    \frac{d\rho_\phi}{dN}
    =-3(1+w_\phi)\rho_\phi,
    \qquad
    \frac{d\rho_{\rm other}}{dN}
    =-3(1+w_{\rm other})\rho_{\rm other}.
\label{eq:separate_density_evolution}
\end{equation}
Differentiating \autoref{eq:scalar_fraction} and using
\autoref{eq:separate_density_evolution}, we obtain
\begin{equation}
    \frac{d\Omega_\phi}{dN}
    =
    3\Omega_\phi(1-\Omega_\phi)
    \left(w_{\rm other}-w_\phi\right).
\label{eq:fraction_evolution}
\end{equation}
The term $3\Omega_\phi(1-\Omega_\phi)$ is always positive. If $w_\phi<w_{\rm other}$, then $\Omega_\phi$ increases with the evolution and if $w_\phi>w_{\rm other}$ it decreases.


A local maximum in $\Omega_\phi$ at $a=a_c$ requires
$w_\phi-w_{\rm other}$ to change from negative to positive. 
In addition, we require
\begin{equation}
    \left.
    \frac{d^2\Omega_\phi}{dN^2}
    \right|_{N_c}
    =
    -3\Omega_{\phi,c}(1-\Omega_{\phi,c})
    \left.
    \frac{d(w_\phi-w_{\rm other})}{dN}
    \right|_{N_c}<0,
\label{eq:peak_second_derivative}
\end{equation}
where $N_c=\ln a_c$.
If the non-scalar components are radiation and
pressureless matter,
\begin{equation}
    w_{\rm other}
    =
    \frac{\rho_r}{3(\rho_r+\rho_m)}.
\label{eq:radiation_matter_eos}
\end{equation}
The equation of state required at the peak is therefore $w_\phi\simeq1/3$ during radiation domination and $w_\phi\simeq0$ during matter domination. At radiation--matter equality, $\rho_r=\rho_m$, and the exact value is $w_\phi=1/6$. Each case also requires a change in sign of $w_\phi-w_{\rm other}$ 
and \autoref{eq:peak_second_derivative}.

The field is initially frozen, so $w_{\phi,\rm ini}=-1$. Thus,  $w_\phi<w_{\rm other}$ and the scalar fraction $\Omega_\phi$ therefore grows. As the expansion rate decreases, the field can roll down the potential and its equation of state can rise above $-1$ \citep{Poulin:2018cxd,Smith:2019ihp}. The scalar fraction continues to grow while $w_\phi<w_{\rm other}$. It reaches a peak when $w_\phi$ crosses $w_{\rm other}$ from below.
After this crossing, the scalar density decreases faster than the combined radiation and matter density, so $\Omega_\phi$ falls. Note that for this  process, it is not necessary at all that the scalar field dominates the total energy density at any stage of this early era.

Thus, in our model, to have a peak, the steep term $V_0x^\alpha$ in \autoref{eq:double_power} is intended to drive this early evolution. The dominance conditions in \autoref{eq:initial_energy_force_dominance} allow this term to control the field's motion.  As the field evolves, $V_0x^\alpha$ must become subdominant to $V_1x^\beta$ for the intended late-time behaviour. The latter term can then support accelerated expansion.

Alternatively, if any of the above conditions is not satisfied, this double power potential can have no early peak, and it can act as a standard quintessence.

\begin{figure}[!htbp]
\centering
\includegraphics[width=0.9\linewidth]{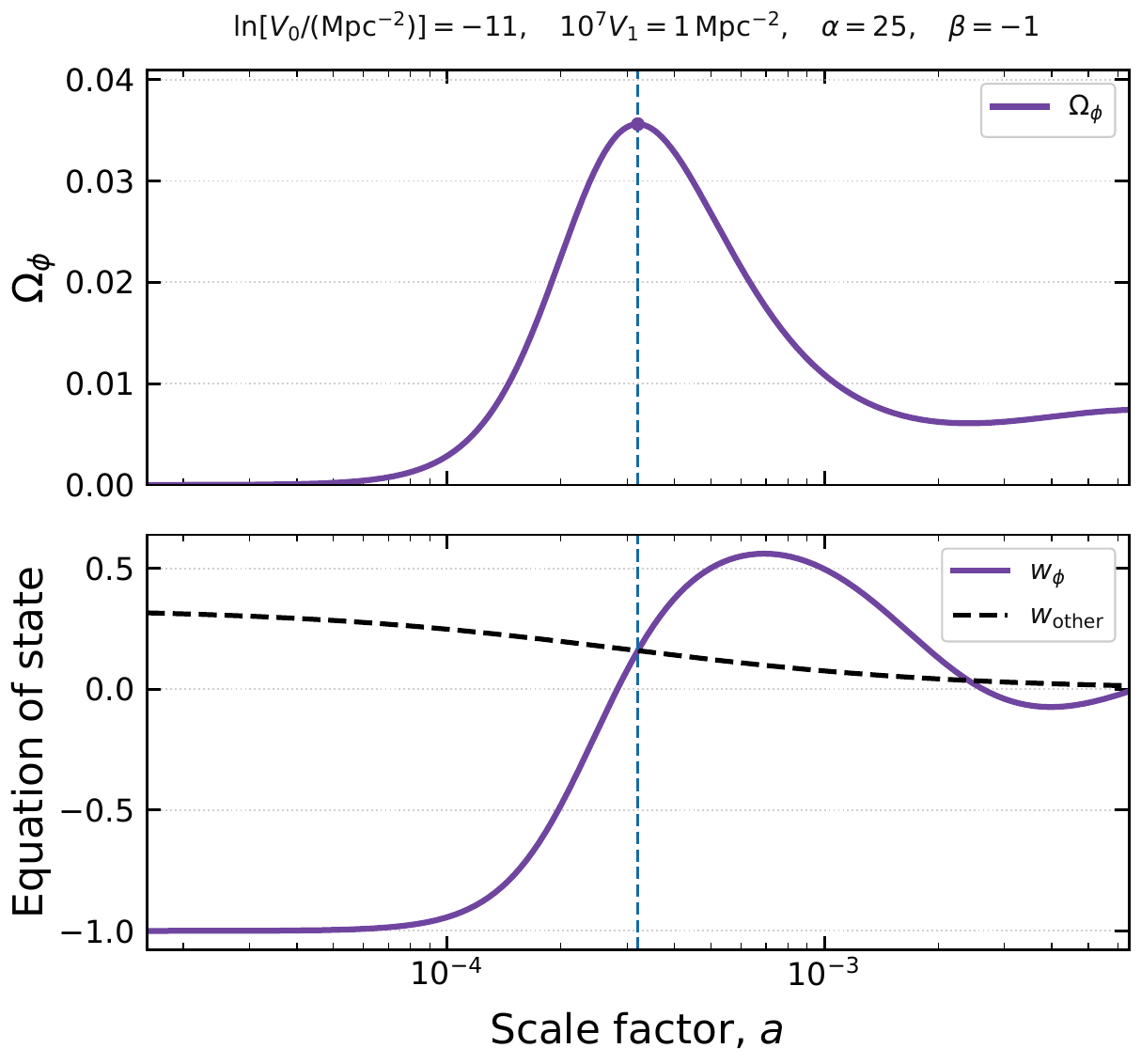}
\caption{The early maximum of the scalar fraction (upper panel) and its relation to the equation-of-state crossing (lower panel). The vertical line marks $a_c\simeq3.2\times10^{-4}$, where $w_\phi$ crosses $w_{\rm other}$. At this maximum, $f_{\rm EDE}\simeq0.036$ and $z_c\simeq3122$.}
\label{fig:early_peak_crossing}
\end{figure}

\begin{figure}[!htbp]
\centering
\includegraphics[width=0.9\linewidth]{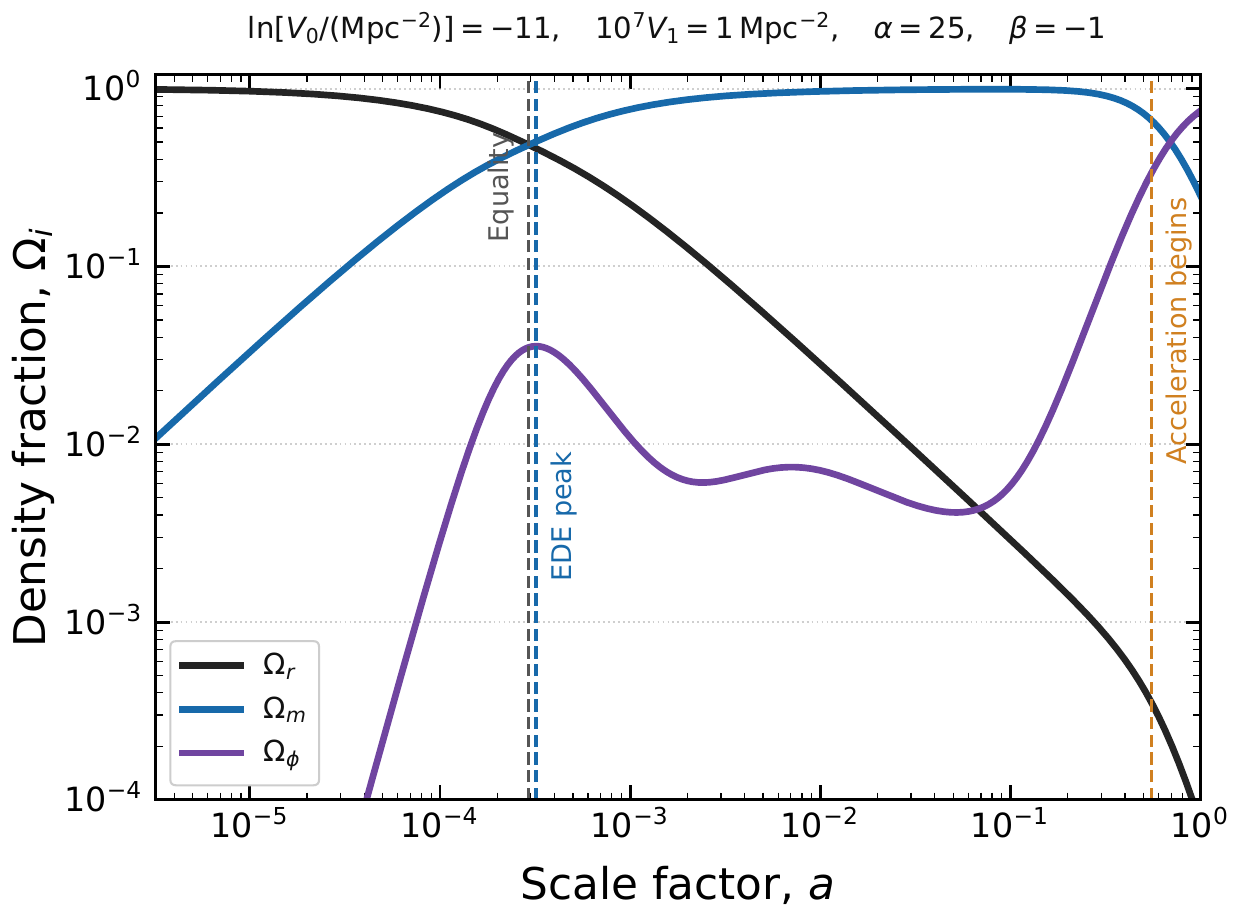}
\caption{Comparative evolution of the radiation, matter, and scalar density fractions. Vertical lines mark radiation--matter equality, the selected EDE maximum, and the onset of acceleration. The scalar fraction declines after the selected maximum and eventually rises to dominate at late times.}
\label{fig:Omega}
\end{figure}


For an early local maximum 
we define
\begin{equation}
    f_{\rm EDE}\equiv\Omega_\phi(a_c),
    \qquad
    z_c\equiv a_c^{-1}-1,
\label{eq:ede_peak_parameters}
\end{equation}
with the present scale factor normalised to unity.
The maximum is selected from the early part of the evolution, before the later rise associated with dark-energy domination. It need not be the largest value of $\Omega_\phi$ over the full history, since the scalar fraction can be larger today.

In \autoref{fig:early_peak_crossing}, the scalar fraction grows while $w_\phi<w_{\rm other}$. It reaches a maximum at $w_\phi=w_{\rm other}$ and then falls while $w_\phi>w_{\rm other}$. The scalar remains subdominant throughout this early era. Finally, we show the entire evolution of the scalar field from an early time to the present compared to other components in \autoref{fig:Omega}.

\section{Observational data}
\label{sec:data}

We constrain the model using CMB temperature, polarization, and lensing measurements, together with DESI DR2 BAO and Pantheon+ supernova data. We also consider the SH0ES calibration of the supernova absolute magnitude. The data sets and likelihoods are as follows.

\begin{itemize}

\item \textbf{Planck low-$\ell$ temperature:} We use the Planck 2018 Commander temperature likelihood over $2\leq\ell\leq29$ \citep{Planck:2019nip}.

\item \textbf{Planck intermediate-$\ell$ spectra:} We use the foreground marginalized Planck 2018 $TT$, $TE$, and $EE$ likelihood \citep{Planck:2019nip}.     
We apply cuts $\ell_{\max}=(1000,600,600)$ respectively, using the \texttt{PlanckActCut} implementation. 

\item \textbf{ACT DR6 temperature and polarization:} We use the foreground-marginalized CMB-only likelihood from ACT Data Release 6 \citep{AtacamaCosmologyTelescope:2025blo}. We select the available $TT$, $TE$, and $EE$ band powers within $600\leq\ell\leq8500$. The ACT temperature-calibration parameter is shared with Planck, $A_{\rm ACT}=A_{\rm Planck}$,     while the ACT polarization-calibration parameter, $p_{\rm ACT}$, is varied independently.    

\item \textbf{SPT-3G D1 temperature and polarization:} We include the first data release from the third-generation camera on the South Pole Telescope, based on observations taken in 2019 and 2020 \citep{SPT-3G:2025bzu}. These measurements cover $400\leq\ell\leq3000$ in $TT$ and $400\leq\ell\leq4000$ in $TE$ and $EE$. We use the foreground-marginalized \texttt{SPT-lite} likelihood through \texttt{candl} \citep{Balkenhol:2024sbv}. We sample the SPT temperature and polarization calibration parameters, $T_{\rm cal}$ and $E_{\rm cal}$. We apply the priors defined in our \texttt{Cobaya} configuration \citep{Torrado:2020dgo} instead of the default priors built into the SPT likelihood.

\item \textbf{Planck and ACT lensing:} We use the ACT DR6 lensing likelihood \citep{ACT:2023dou,ACT:2023kun} with its \texttt{actplanck\_baseline} setting. This combines ACT DR6 lensing with the Planck PR4 lensing reconstruction \citep{Carron:2022eyg}. We retain the likelihood corrections required when combining lensing with primary CMB spectra by setting \texttt{lens\_only=false}.    

\item \textbf{SPT-3G MUSE lensing:} We include the CMB lensing reconstruction obtained from the 2019--2020 SPT-3G polarization observations using the Marginal Unbiased Score Expansion (MUSE) method \citep{SPT-3G:2024atg}. We use only the lensing-potential power-spectrum component of this likelihood. Its delensed $EE$ component is not included.    

\item \textbf{DESI DR2 BAO:}
We use the BAO distance measurements from the first three years of DESI observations \citep{DESI:2025zgx}. These include galaxy, quasar, and Lyman-$\alpha$ forest measurements. We use the public \texttt{bao.desi\_dr2} likelihood with its supplied covariance matrix. The measured distances are expressed relative to the sound horizon at the baryon-drag epoch, $r_d$, which is calculated self-consistently within our model. 

\item \textbf{Pantheon+ supernovae:} We use the Pantheon+ compilation of standardized Type Ia supernova light curves \citep{Scolnic:2021amr,Brout:2022vxf}. The full compilation contains 1701 light curves from 1550 distinct supernovae and extends to $z\simeq2.26$. We adopt the selection and statistical and systematic covariance supplied with the corresponding public likelihood.    

\item \textbf{SH0ES calibration:} For  calibrated supernova analysis, we use \texttt{sn.pantheonplusshoes}. This likelihood includes the SH0ES Cepheid distances to supernova host galaxies \citep{Riess:2021jrx}, which calibrate the supernova absolute magnitude $M_B$ that is varied in our analysis. 
We do not add a separate Gaussian SH0ES likelihood for $H_0$.

\end{itemize}

In this paper, CMB denotes the combined Planck, ACT, and SPT temperature and polarization likelihoods, together with the lensing likelihoods listed above. This combination does not include Planck low-multipole polarization or BICEP--Keck data. PPSHOES denotes the SH0ES-calibrated Pantheon+ sample. Our analysis uses the combined dataset CMB+\allowbreak DESI+\allowbreak PPSHOES.

\section{Model parameters and priors}
\label{sec:parameters}


We fix the scalar potential powers to $\alpha=25$ and $\beta=-1$ and the initial field values to $\phi_{\rm ini}=2M_{\rm Pl}$ and $\phi'_{\rm ini}=0$. The two potential amplitudes $V_0, V_1$ remain free, so the sampled parameters are
\begin{equation}
    \begin{aligned}
        \boldsymbol{\Theta}=\Big\{
        \omega_b,\,
        \omega_{\rm cdm},\,
        \ln(10^{10}A_s),\,
        n_s,\,
        \tau_{\rm reio},\,
        u_0,\,
        u_1,
        M_B,\,
        A_{\rm Planck},\,
        p_{\rm ACT},\,
        T_{\rm cal},\,
        E_{\rm cal}
        \Big\}.
    \end{aligned}
    \label{eq:sampled_parameters}
\end{equation}
The first five are standard cosmological parameters, with the next two being dimensionless scalar parameters:
\begin{equation}
    u_0
    \equiv
    \ln\left[
        \frac{V_0}{\mathrm{Mpc}^{-2}}
    \right],
    \qquad
    u_1
    \equiv
    \frac{10^7V_1}{\mathrm{Mpc}^{-2}} \,.
\label{eq:sampled_potential_amplitudes}
\end{equation}
The remaining five are the supernova absolute magnitude $M_B$ and the CMB calibrations. The ACT temperature calibration is not an independent parameter, since $A_{\rm ACT}=A_{\rm Planck}$. 

With $\alpha$, $\beta$, and the initial field conditions fixed, our model has seven free cosmological parameters, while flat $\Lambda$CDM has six. The two potential amplitudes replace the single parameter that fixes the present dark-energy density. Our model therefore introduces only one additional parameter.

\begin{table}[!htbp]
\centering
\begin{tabular}{lc}
\hline
Parameter & Prior \\
\hline
$\omega_b$         & $\mathcal{U}(0.005,\,0.1)$ \\
$\omega_{\rm cdm}$ & $\mathcal{U}(0.001,\,0.99)$ \\
$\ln(10^{10}A_s)$  & $\mathcal{U}(1.61,\,3.91)$ \\
$n_s$             & $\mathcal{U}(0.8,\,1.2)$ \\
$\tau_{\rm reio}$  & $\mathcal{N}(0.051,\,0.006^2)$ \\
$u_0$             & $\mathcal{U}(-100,\,0)$ \\
$u_1$             & $\mathcal{U}(0.01,\,3)$ \\
$M_B$             & $\mathcal{U}(-21,\,-18)$ \\
$A_{\rm Planck}$  & $\mathcal{U}(0.5,\,1.5)$ \\
$p_{\rm ACT}$     & $\mathcal{U}(0.9,\,1.1)$ \\
$T_{\rm cal}$     & $\mathcal{N}(1,\,0.0036^2)$ \\
$E_{\rm cal}$     & $\mathcal{U}(0.8,\,1.2)$ \\
\hline
\end{tabular}
\caption{
Priors on the sampled parameters. Here, $\mathcal{U}(a,b)$ denotes a uniform distribution over $[a,b]$, and $\mathcal{N}(\mu,\sigma^2)$ denotes a Gaussian distribution with mean $\mu$ and variance $\sigma^2$. The potential variables $u_0$ and $u_1$ are defined in \autoref{eq:sampled_potential_amplitudes}.
}
\label{tab:parameter_priors}
\end{table}

We adopt the priors listed in \autoref{tab:parameter_priors}. All priors are uniform except those on $\tau_{\rm reio}$ and $T_{\rm cal}$, which are Gaussian. For a Gaussian prior, we quote its mean and standard deviation.

\section{Results}
\label{sec:results}

\begin{table}[!htbp]
\centering
\begin{tabular}{lc}
\hline
Parameter & CMB+DESI+PPSHOES \\
\hline
$\ln(10^{10}A_s)$ & $3.077\pm0.017$ \\
$n_s$ & $0.9779\pm0.0030$ \\
$\tau_{\rm reio}$ & $0.062\pm0.012$ \\
$\omega_b$ & $0.022611\pm0.000094$ \\
$\omega_{\rm cdm}$ & $0.12313\pm0.00065$ \\
$u_0$ & $-11.34\pm0.10$ \\
$u_1$ & $0.788\pm0.010$ \\
$M_B$ & $-19.3727\pm0.0078$ \\
$A_{\rm Planck}$ & $1.0072\pm0.0050$ \\
$p_{\rm ACT}$ & $1.0040\pm0.0025$ \\
$T_{\rm cal}$ & $1.0030\pm0.0056$ \\
$E_{\rm cal}$ & $1.0099\pm0.0042$ \\
\hline
\end{tabular}
\caption{
Parameter estimates and reported $1\sigma$ uncertainties for the double-power scalar-field model using CMB+DESI+PPSHOES. The potential variables $u_i$ are defined in \autoref{eq:sampled_potential_amplitudes}.
}
\label{tab:parameter_constraints}
\end{table}

\begin{figure}[!htbp]
\centering
\includegraphics[width=0.9\linewidth]{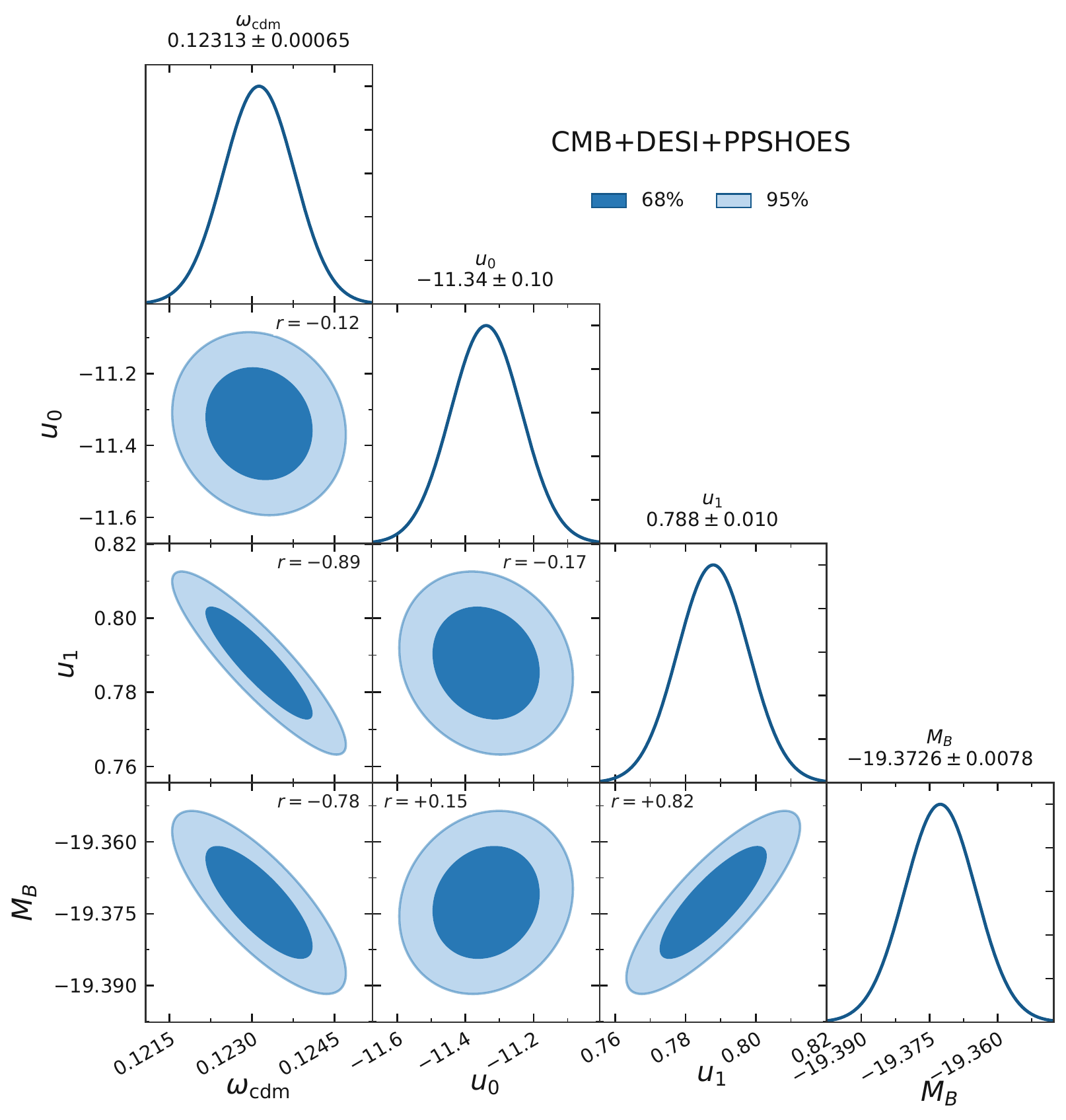}
\caption{
Constraints on $\omega_{\rm cdm}$, $u_0$, $u_1$, and $M_B$ from CMB+DESI+PPSHOES. The potential variables $u_0$ and $u_1$ are defined in \autoref{eq:sampled_potential_amplitudes}.
}
\label{fig:triangle_ppshoes}
\end{figure}

We present the results for CMB+\allowbreak DESI+\allowbreak PPSHOES. 
We focus on four parameters $\omega_{\rm cdm}$, $u_0$, $u_1$, and $M_B$ in the main triangle plot in
\autoref{fig:triangle_ppshoes}, but full constraints are listed in \autoref{tab:parameter_constraints}.

The potential amplitudes are
\begin{equation}
    u_0=-11.34\pm0.10,
    \qquad
    u_1=0.788\pm0.010\,,
\label{eq:potential_constraints_ppshoes}
\end{equation}
which place the initial field deep in the regime dominated by $V_0x^\alpha$. At the best-fit point, the initial ratios are
\begin{equation}
    \mathcal R_{\rm ini}\simeq1.01\times10^{10},
    \qquad
    |\alpha/\beta|\,\mathcal R_{\rm ini}
    \simeq2.53\times10^{11}.
    \label{eq:initial_ratio_ppshoes}
\end{equation}
The steep term therefore dominates both the initial potential and the magnitude of its slope. As the field evolves towards smaller values, $\mathcal R$ decreases.
At the present epoch, 
\begin{equation}
    x_0\simeq0.7,
    \qquad
    \mathcal R_0\simeq0.04.
    \label{eq:present_ratio_ppshoes}
\end{equation}
Thus, the inverse-power term contributes approximately $96\%$ of the total potential today. This solution exhibits the intended transition from early dominance of the steep term to late dominance
of the inverse-power term.

\subsection{Implications for
the Hubble and clustering tensions}
\label{subsec:parameter_comparison}

For CMB+DESI+PPSHOES, the scalar-field model favours higher baryon and cold-dark-matter densities, primordial amplitude, and spectral index than flat $\Lambda$CDM. It gives $M_B=-19.3727\pm0.0078$, compared with $M_B=-19.4072\pm0.0077$ in $\Lambda$CDM. These differences reflect the joint constraints on the expansion history and supernova calibration.
%
\begin{table}[!htbp]
\centering
\begin{tabular}{lcc}
\hline
& \multicolumn{2}{c}{CMB+DESI+PPSHOES} \\
Parameter & Scalar field & $\Lambda$CDM \\
\hline
$\Omega_{\rm m0}$ & $0.3028\pm0.0034$ & $0.3008\pm0.0032$ \\
$H_0$ & $69.53\pm0.26$ & $68.45\pm0.24$ \\
$\sigma_8$ & $0.8010\pm0.0064$ & $0.8115\pm0.0040$ \\
$S_8$ & $0.8047\pm0.0084$ & $0.8126\pm0.0063$ \\
$r_d$ & $144.56\pm0.18$ & $147.47\pm0.18$ \\
$100\theta_*$ & $1.04430\pm0.00072$ & $1.04449\pm0.00041$ \\
$r_s(z_*)$ & $142.48\pm0.21$ & $145.30\pm0.18$ \\
$D_M(z_*)$ & $13643\pm16$ & $13911\pm16$ \\
\hline
\end{tabular}
\caption{
Derived cosmological parameters for the scalar-field model and flat $\Lambda$CDM, using the dataset CMB+DESI+PPSHOES. $H_0$ is in ${\rm km\,s^{-1}\,Mpc^{-1}}$, and distances are in ${\rm Mpc}$.
}
\label{tab:derived_parameters}
\end{table}
%
\autoref{tab:derived_parameters} compares the derived cosmological quantities in the scalar-field model and flat $\Lambda$CDM. 

\paragraph{Expansion rate and sound horizons.}

The scalar-field model gives
\begin{equation}
    H_0=69.53\pm0.26\,
    {\rm km\,s^{-1}\,Mpc^{-1}},
\label{eq:H0_scalar_ppshoes}
\end{equation}
compared with $68.45\pm0.24\,{\rm km\,s^{-1}\,Mpc^{-1}}$ in $\Lambda$CDM. The scalar-model central value is higher by approximately $1.09\,{\rm km\,s^{-1}\,Mpc^{-1}}$. The SH0ES distance ladder gives $H_0=73.04\pm1.04\, {\rm km\,s^{-1}\,Mpc^{-1}}$ \citep{Riess:2021jrx}. The separation from its central value decreases from approximately $4.59\,{\rm km\,s^{-1}\,Mpc^{-1}}$ in $\Lambda$CDM to $3.51\,{\rm km\,s^{-1}\,Mpc^{-1}}$ in the scalar-field model. The inferred expansion rate therefore moves towards SH0ES, although a substantial separation remains.

The higher inferred expansion rate accompanies smaller sound horizons. The scalar-field model gives $r_d=144.56\pm0.18\,{\rm Mpc}$, compared with $147.47\pm0.18\,{\rm Mpc}$ in $\Lambda$CDM. The sound horizon at last scattering decreases from $145.30\pm0.18\,{\rm Mpc}$ to $142.48\pm0.21\,{\rm Mpc}$ compared to $\Lambda$CDM. The distance to last scattering also decreases, from $13911\pm16\,{\rm Mpc}$ to $13643\pm16\,{\rm Mpc}$. Thus, $r_s(z_*)$ and $D_M(z_*)$ both decrease by approximately $1.9\%$. Their nearly proportional changes leave $\theta_*=r_s(z_*)/D_M(z_*)$ almost unchanged. This illustrates the geometrical mechanism of EDE. 

\paragraph{Matter clustering.}

The clustering amplitude is probed via $S_8$, defined in \autoref{eq:S8_definition}, which depends on both $\sigma_8$ and $\Omega_{m0}$. For CMB+DESI+PPSHOES, the scalar-field model gives
\begin{equation}
    \sigma_8=0.8010\pm0.0064,
    \qquad
    S_8=0.8047\pm0.0084,
    \label{eq:clustering_scalar_ppshoes}
\end{equation}
compared with $\sigma_8=0.8115\pm0.0040$ and $S_8=0.8126\pm0.0063$ in flat $\Lambda$CDM. The central values are lower by approximately $1.3\%$ and $1.0\%$, respectively.

For comparison, the DES Year~6 analysis gives $S_8=0.789\pm0.012$ within flat $\Lambda$CDM \citep{DES:2026fyc}. The scalar-field central value lies closer to this measurement. This suggests improved agreement with DES, although a quantitative assessment requires testing the scalar-field predictions against the DES likelihood, which we leave for future study.

\subsection{Comparison with flat $\Lambda$CDM}
\label{subsec:best_fit_results}

\begin{table}[!htbp]
\centering
\begin{tabular}{lc}
\hline
Quantity & CMB+DESI+PPSHOES \\
\hline
$\Delta\chi^2_{\rm data}$ & $-6.895$ \\
$\Delta\chi^2_{\rm Gaussian\ priors}$ & $+3.138$ \\
$\Delta\chi^2_{\rm objective}$ & $-3.757$ \\
\hline
\end{tabular}
\caption{
Differences $\Delta\chi^2\equiv\chi^2_{\rm scalar}-\chi^2_{\Lambda{\rm CDM}}$ at the optimized parameter points for CMB+\allowbreak DESI+\allowbreak PPSHOES. The objective is the sum of the data $\chi^2$ and Gaussian parameter-prior penalties. Negative values indicate a lower $\chi^2$ for the scalar-field model.
}
\label{tab:model_comparison}
\end{table}

We compare the scalar-field model with flat $\Lambda$CDM using the same CMB+DESI+\allowbreak PPSHOES likelihoods and normalization conventions. The optimization minimizes the sum of the data $\chi^2$ and Gaussian parameter-prior penalties. Their differences are listed separately in \autoref{tab:model_comparison}.
At the optimized points, the scalar-field model improves the data fit by $\Delta\chi^2_{\rm data}=-6.895$. Its Gaussian-prior penalty is larger by $3.138$, partly offsetting this improvement. The total objective is therefore lower by $\Delta\chi^2_{\rm objective}=-3.757$.

With the potential powers and initial field conditions fixed, the scalar-field model has one additional free parameter relative to flat $\Lambda$CDM. For data-only maximum-likelihood fits, the difference in the Akaike information criterion is
\begin{equation}
    \Delta{\rm AIC}
    \equiv{\rm AIC}_{\rm scalar}-{\rm AIC}_{\Lambda{\rm CDM}}=
    \Delta\chi^2_{{\rm data},\min}+2
    = -4.895\,.
    \label{eq:results_delta_aic}
\end{equation}
The quoted value assumes that the optimized data $\chi^2$ values are also the data-only minima.
Under this assumption, \autoref{eq:results_delta_aic}
shows that the improvement in the data fit outweighs the penalty for the additional parameter.

\newpage
\section{Conclusions}
\label{sec:conclusions}

We investigated whether a single canonical scalar field can provide an early dark-energy contribution and also drive late-time cosmic acceleration. We considered a  potential with double powers, $\alpha=25$ and $\beta=-1$, with initial value $\phi_{\rm ini}=2M_{\rm Pl}$ and zero initial velocity. The steep positive-power term controls the early evolution, while the inverse-power term supports late-time acceleration. Both eras arise from the same potential and scalar-field trajectory.

We tested the model against CMB measurements, DESI DR2 BAO distances, and Pantheon+ supernovae with the SH0ES host-distance calibration. At the best-fit point, the steep term dominates the initial potential and its slope. The scalar energy-density fraction subsequently reaches an early peak and decreases before rising again at late times. The inverse-power term contributes approximately $96\%$ of the potential today.

For CMB+DESI+PPSHOES, the scalar-field model gives $H_0=69.53\pm0.26\,{\rm km\,s^{-1}\,Mpc^{-1}}$, compared with $68.45\pm0.24\,{\rm km\,s^{-1}\,Mpc^{-1}}$ in flat $\Lambda$CDM. Its sound horizon and comoving distance to last scattering are both approximately $1.9\%$ smaller, leaving the angular acoustic scale almost unchanged. The inferred $H_0$ moves towards the SH0ES measurement, although a substantial separation remains.

The scalar-field model also gives a lower clustering amplitude, with $S_8=0.8047\pm0.0084$, compared with $0.8126\pm0.0063$ in flat $\Lambda$CDM. This shift brings the central value closer to the DES Year~6 result. Establishing whether the model reduces the clustering tension requires a consistent analysis of weak-lensing data within the scalar-field model.

The data fit improves by $\Delta\chi^2_{\rm data}=-6.90$ relative to flat $\Lambda$CDM. Including Gaussian parameter-prior penalties reduces the improvement to $\Delta\chi^2_{\rm objective}=-3.76$. With the potential powers and initial field conditions fixed, the model introduces only one additional free parameter. These results demonstrate that a single minimally coupled scalar field can produce an early energy-density peak and late-time acceleration while improving the fit to CMB+DESI+PPSHOES.

\acknowledgments
The authors acknowledge the Centre for High Performance Computing (CHPC), South Africa, for providing computational resources to this research project. {BRD is supported by the South African Radio Astronomy Observatory and the National Research Foundation.}

\clearpage
\bibliographystyle{JHEP}
\bibliography{references}

@article{SupernovaSearchTeam:1998fmf,
    author = "Riess, Adam G. and others",
    collaboration = "Supernova Search Team",
    title = "{Observational evidence from supernovae for an accelerating universe and a cosmological constant}",
    eprint = "astro-ph/9805201",
    archivePrefix = "arXiv",
    doi = "10.1086/300499",
    journal = "Astron. J.",
    volume = "116",
    pages = "1009--1038",
    year = "1998"
}

@article{SupernovaCosmologyProject:1998vns,
    author = "Perlmutter, S. and others",
    collaboration = "Supernova Cosmology Project",
    title = "{Measurements of $\Omega$ and $\Lambda$ from 42 high redshift supernovae}",
    eprint = "astro-ph/9812133",
    archivePrefix = "arXiv",
    reportNumber = "LBNL-41801, LBL-41801",
    doi = "10.1086/307221",
    journal = "Astrophys. J.",
    volume = "517",
    pages = "565--586",
    year = "1999"
}

@article{BOSS:2016wmc,
    author = "Alam, Shadab and others",
    collaboration = "BOSS",
    title = "{The clustering of galaxies in the completed SDSS-III Baryon Oscillation Spectroscopic Survey: cosmological analysis of the DR12 galaxy sample}",
    eprint = "1607.03155",
    archivePrefix = "arXiv",
    primaryClass = "astro-ph.CO",
    doi = "10.1093/mnras/stx721",
    journal = "Mon. Not. Roy. Astron. Soc.",
    volume = "470",
    number = "3",
    pages = "2617--2652",
    year = "2017"
}

@article{eBOSS:2020yzd,
    author = "Alam, Shadab and others",
    collaboration = "eBOSS",
    title = "{Completed SDSS-IV extended Baryon Oscillation Spectroscopic Survey: Cosmological implications from two decades of spectroscopic surveys at the Apache Point Observatory}",
    eprint = "2007.08991",
    archivePrefix = "arXiv",
    primaryClass = "astro-ph.CO",
    doi = "10.1103/PhysRevD.103.083533",
    journal = "Phys. Rev. D",
    volume = "103",
    number = "8",
    pages = "083533",
    year = "2021"
}

@article{Peebles:2002gy,
    author = "Peebles, P. J. E. and Ratra, Bharat",
    editor = "Hsu, Jong-Ping and Fine, D.",
    title = "{The Cosmological Constant and Dark Energy}",
    eprint = "astro-ph/0207347",
    archivePrefix = "arXiv",
    reportNumber = "KSUPT-02-3",
    doi = "10.1103/RevModPhys.75.559",
    journal = "Rev. Mod. Phys.",
    volume = "75",
    pages = "559--606",
    year = "2003"
}

@article{Carroll:2000fy,
    author = "Carroll, Sean M.",
    title = "{The Cosmological constant}",
    eprint = "astro-ph/0004075",
    archivePrefix = "arXiv",
    reportNumber = "EFI-2000-13",
    doi = "10.12942/lrr-2001-1",
    journal = "Living Rev. Rel.",
    volume = "4",
    pages = "1",
    year = "2001"
}

@article{Velten:2014nra,
    author = "Velten, H.E.S. and vom Marttens, R.F. and Zimdahl, W.",
    title = "{Aspects of the cosmological \textquotedblleft{}coincidence problem\textquotedblright{}}",
    eprint = "1410.2509",
    archivePrefix = "arXiv",
    primaryClass = "astro-ph.CO",
    doi = "10.1140/epjc/s10052-014-3160-4",
    journal = "Eur. Phys. J. C",
    volume = "74",
    number = "11",
    pages = "3160",
    year = "2014"
}

@article{DiValentino:2021izs,
    author = "Di Valentino, Eleonora and Mena, Olga and Pan, Supriya and Visinelli, Luca and Yang, Weiqiang and Melchiorri, Alessandro and Mota, David F. and Riess, Adam G. and Silk, Joseph",
    title = "{In the realm of the Hubble tension\textemdash{}a review of solutions}",
    eprint = "2103.01183",
    archivePrefix = "arXiv",
    primaryClass = "astro-ph.CO",
    reportNumber = "IPPP/20/108",
    doi = "10.1088/1361-6382/ac086d",
    journal = "Class. Quant. Grav.",
    volume = "38",
    number = "15",
    pages = "153001",
    year = "2021"
}

@article{Brout:2022vxf,
    author = "Brout, Dillon and others",
    title = "{The Pantheon+ Analysis: Cosmological Constraints}",
    eprint = "2202.04077",
    archivePrefix = "arXiv",
    primaryClass = "astro-ph.CO",
    doi = "10.3847/1538-4357/ac8e04",
    journal = "Astrophys. J.",
    volume = "938",
    number = "2",
    pages = "110",
    year = "2022"
}

@article{Hogg:1999ad,
    author = "Hogg, David W.",
    title = "{Distance measures in cosmology}",
    eprint = "astro-ph/9905116",
    archivePrefix = "arXiv",
    month = "5",
    year = "1999"
}

@article{Hu:1995en,
    author = "Hu, Wayne and Sugiyama, Naoshi",
    title = "{Small scale cosmological perturbations: An Analytic approach}",
    eprint = "astro-ph/9510117",
    archivePrefix = "arXiv",
    reportNumber = "IASSNS-AST-95-42, CFPA-TH-95-18, UTAP-212",
    doi = "10.1086/177989",
    journal = "Astrophys. J.",
    volume = "471",
    pages = "542--570",
    year = "1996"
}

@article{Scolnic:2021amr,
    author = "Scolnic, Dan and others",
    title = "{The Pantheon+ Analysis: The Full Data Set and Light-curve Release}",
    eprint = "2112.03863",
    archivePrefix = "arXiv",
    primaryClass = "astro-ph.CO",
    doi = "10.3847/1538-4357/ac8b7a",
    journal = "Astrophys. J.",
    volume = "938",
    number = "2",
    pages = "113",
    year = "2022"
}

@article{Riess:2021jrx,
    author = "Riess, Adam G. and others",
    title = "{A Comprehensive Measurement of the Local Value of the Hubble Constant with 1 km/s/Mpc Uncertainty from the Hubble Space Telescope and the SH0ES Team}",
    eprint = "2112.04510",
    archivePrefix = "arXiv",
    primaryClass = "astro-ph.CO",
    doi = "10.3847/2041-8213/ac5c5b",
    journal = "Astrophys. J. Lett.",
    volume = "934",
    number = "1",
    pages = "L7",
    year = "2022"
}

@article{ACT:2023kun,
    author = "Madhavacheril, Mathew S. and others",
    collaboration = "ACT",
    title = "{The Atacama Cosmology Telescope: DR6 Gravitational Lensing Map and Cosmological Parameters}",
    eprint = "2304.05203",
    archivePrefix = "arXiv",
    primaryClass = "astro-ph.CO",
    reportNumber = "FERMILAB-PUB-23-206-PPD",
    doi = "10.3847/1538-4357/acff5f",
    journal = "Astrophys. J.",
    volume = "962",
    number = "2",
    pages = "113",
    year = "2024"
}

@article{ACT:2023dou,
    author = "Qu, Frank J. and others",
    collaboration = "ACT",
    title = "{The Atacama Cosmology Telescope: A Measurement of the DR6 CMB Lensing Power Spectrum and Its Implications for Structure Growth}",
    eprint = "2304.05202",
    archivePrefix = "arXiv",
    primaryClass = "astro-ph.CO",
    reportNumber = "FERMILAB-PUB-23-237-PPD, FERMILAB-PUB-23-237-PPD",
    doi = "10.3847/1538-4357/acfe06",
    journal = "Astrophys. J.",
    volume = "962",
    number = "2",
    pages = "112",
    year = "2024"
}

@article{Carron:2022eyg,
    author = "Carron, Julien and Mirmelstein, Mark and Lewis, Antony",
    title = "{CMB lensing from Planck PR4~maps}",
    eprint = "2206.07773",
    archivePrefix = "arXiv",
    primaryClass = "astro-ph.CO",
    doi = "10.1088/1475-7516/2022/09/039",
    journal = "JCAP",
    volume = "09",
    pages = "039",
    year = "2022"
}

@article{DESI:2025zgx,
    author = "Abdul Karim, M. and others",
    collaboration = "DESI",
    title = "{DESI DR2 results. II. Measurements of baryon acoustic oscillations and cosmological constraints}",
    eprint = "2503.14738",
    archivePrefix = "arXiv",
    primaryClass = "astro-ph.CO",
    reportNumber = "FERMILAB-PUB-25-0169-PPD",
    doi = "10.1103/tr6y-kpc6",
    journal = "Phys. Rev. D",
    volume = "112",
    number = "8",
    pages = "083515",
    year = "2025"
}

@article{Poulin:2023lkg,
    author = "Poulin, Vivian and Smith, Tristan L. and Karwal, Tanvi",
    title = "{The Ups and Downs of Early Dark Energy solutions to the Hubble tension: A review of models, hints and constraints circa 2023}",
    eprint = "2302.09032",
    archivePrefix = "arXiv",
    primaryClass = "astro-ph.CO",
    doi = "10.1016/j.dark.2023.101348",
    journal = "Phys. Dark Univ.",
    volume = "42",
    pages = "101348",
    year = "2023"
}

@article{Pantos:2026koc,
    author = "Pantos, Ioannis and Perivolaropoulos, Leandros",
    title = "{Status of the $S_8$ Tension: A 2026 Review of Probe Discrepancies}",
    eprint = "2602.12238",
    archivePrefix = "arXiv",
    primaryClass = "astro-ph.CO",
    doi = "10.1016/j.dark.2026.102286",
    journal = "Phys. Dark Univ.",
    volume = "52",
    pages = "102286",
    year = "2026"
}

@article{Poulin:2025nfb,
    author = "Poulin, Vivian and Smith, Tristan L. and Calder{\'o}n, Rodrigo and Simon, Th{\'e}o",
    title = "{Impact of ACT DR6 and DESI DR2 for early dark energy and the Hubble tension}",
    eprint = "2505.08051",
    archivePrefix = "arXiv",
    primaryClass = "astro-ph.CO",
    doi = "10.1103/bx25-1g5d",
    journal = "Phys. Rev. D",
    volume = "113",
    number = "6",
    pages = "063519",
    year = "2026"
}

@article{Weinberg:1988cp,
    author = "Weinberg, Steven",
    editor = "Hsu, Jong-Ping and Fine, D.",
    title = "{The Cosmological Constant Problem}",
    reportNumber = "UTTG-12-88",
    doi = "10.1103/RevModPhys.61.1",
    journal = "Rev. Mod. Phys.",
    volume = "61",
    pages = "1--23",
    year = "1989"
}

@article{Planck:2018vyg,
    author = "Aghanim, N. and others",
    collaboration = "Planck",
    title = "{Planck 2018 results. VI. Cosmological parameters}",
    eprint = "1807.06209",
    archivePrefix = "arXiv",
    primaryClass = "astro-ph.CO",
    doi = "10.1051/0004-6361/201833910",
    journal = "Astron. Astrophys.",
    volume = "641",
    pages = "A6",
    year = "2020",
    note = "[Erratum: Astron.Astrophys. 652, C4 (2021)]"
}

@article{Freedman:2021ahq,
    author = "Freedman, Wendy L.",
    title = "{Measurements of the Hubble Constant: Tensions in Perspective}",
    eprint = "2106.15656",
    archivePrefix = "arXiv",
    primaryClass = "astro-ph.CO",
    doi = "10.3847/1538-4357/ac0e95",
    journal = "Astrophys. J.",
    volume = "919",
    number = "1",
    pages = "16",
    year = "2021"
}

@article{Riess:2024vfa,
    author = "Riess, Adam G. and others",
    title = "{JWST Validates HST Distance Measurements: Selection of Supernova Subsample Explains Differences in JWST Estimates of Local H $_{0}$}",
    eprint = "2408.11770",
    archivePrefix = "arXiv",
    primaryClass = "astro-ph.CO",
    doi = "10.3847/1538-4357/ad8c21",
    journal = "Astrophys. J.",
    volume = "977",
    number = "1",
    pages = "120",
    year = "2024"
}

@article{Heymans:2020gsg,
    author = "Heymans, Catherine and others",
    title = "{KiDS-1000 Cosmology: Multi-probe weak gravitational lensing and spectroscopic galaxy clustering constraints}",
    eprint = "2007.15632",
    archivePrefix = "arXiv",
    primaryClass = "astro-ph.CO",
    doi = "10.1051/0004-6361/202039063",
    journal = "Astron. Astrophys.",
    volume = "646",
    pages = "A140",
    year = "2021"
}

@article{DES:2021wwk,
    author = "Abbott, T. M. C. and others",
    collaboration = "DES",
    title = "{Dark Energy Survey Year 3 results: Cosmological constraints from galaxy clustering and weak lensing}",
    eprint = "2105.13549",
    archivePrefix = "arXiv",
    primaryClass = "astro-ph.CO",
    reportNumber = "FERMILAB-PUB-21-221-AE, DES-2020-0617",
    doi = "10.1103/PhysRevD.105.023520",
    journal = "Phys. Rev. D",
    volume = "105",
    number = "2",
    pages = "023520",
    year = "2022"
}

@article{Li:2023tui,
    author = "Li, Xiangchong and others",
    title = "{Hyper Suprime-Cam Year 3 results: Cosmology from cosmic shear two-point correlation functions}",
    eprint = "2304.00702",
    archivePrefix = "arXiv",
    primaryClass = "astro-ph.CO",
    doi = "10.1103/PhysRevD.108.123518",
    journal = "Phys. Rev. D",
    volume = "108",
    number = "12",
    pages = "123518",
    year = "2023"
}

@article{Wright:2025xka,
    author = "Wright, Angus H. and others",
    title = "{KiDS-Legacy: Cosmological constraints from cosmic shear with the complete Kilo-Degree Survey}",
    eprint = "2503.19441",
    archivePrefix = "arXiv",
    primaryClass = "astro-ph.CO",
    doi = "10.1051/0004-6361/202554908",
    journal = "Astron. Astrophys.",
    volume = "703",
    pages = "A158",
    year = "2025"
}

@article{Knox:2019rjx,
    author = "Knox, Lloyd and Millea, Marius",
    title = "{Hubble constant hunter{\textquoteright}s guide}",
    eprint = "1908.03663",
    archivePrefix = "arXiv",
    primaryClass = "astro-ph.CO",
    doi = "10.1103/PhysRevD.101.043533",
    journal = "Phys. Rev. D",
    volume = "101",
    number = "4",
    pages = "043533",
    year = "2020"
}

@article{Bernal:2016gxb,
    author = "Bernal, Jose Luis and Verde, Licia and Riess, Adam G.",
    title = "{The trouble with $H_0$}",
    eprint = "1607.05617",
    archivePrefix = "arXiv",
    primaryClass = "astro-ph.CO",
    doi = "10.1088/1475-7516/2016/10/019",
    journal = "JCAP",
    volume = "10",
    pages = "019",
    year = "2016"
}

@article{Aylor:2018drw,
    author = "Aylor, Kevin and Joy, MacKenzie and Knox, Lloyd and Millea, Marius and Raghunathan, Srinivasan and Wu, W. L. Kimmy",
    title = "{Sounds Discordant: Classical Distance Ladder {\&} $\Lambda$CDM -based Determinations of the Cosmological Sound Horizon}",
    eprint = "1811.00537",
    archivePrefix = "arXiv",
    primaryClass = "astro-ph.CO",
    doi = "10.3847/1538-4357/ab0898",
    journal = "Astrophys. J.",
    volume = "874",
    number = "1",
    pages = "4",
    year = "2019"
}

@article{Schoneberg:2021qvd,
    author = {Sch{\"o}neberg, Nils and Franco Abell{\'a}n, Guillermo and P{\'e}rez S{\'a}nchez, Andrea and Witte, Samuel J. and Poulin, Vivian and Lesgourgues, Julien},
    title = "{The H0 Olympics: A fair ranking of proposed models}",
    eprint = "2107.10291",
    archivePrefix = "arXiv",
    primaryClass = "astro-ph.CO",
    doi = "10.1016/j.physrep.2022.07.001",
    journal = "Phys. Rept.",
    volume = "984",
    pages = "1--55",
    year = "2022"
}

@article{DiValentino:2020vvd,
    author = "Di Valentino, Eleonora and others",
    title = "{Cosmology Intertwined III: $f \sigma_8$ and $S_8$}",
    eprint = "2008.11285",
    archivePrefix = "arXiv",
    primaryClass = "astro-ph.CO",
    reportNumber = "FERMILAB-PUB-20-495-AE",
    doi = "10.1016/j.astropartphys.2021.102604",
    journal = "Astropart. Phys.",
    volume = "131",
    pages = "102604",
    year = "2021"
}

@article{Poulin:2018cxd,
    author = "Poulin, Vivian and Smith, Tristan L. and Karwal, Tanvi and Kamionkowski, Marc",
    title = "{Early Dark Energy Can Resolve The Hubble Tension}",
    eprint = "1811.04083",
    archivePrefix = "arXiv",
    primaryClass = "astro-ph.CO",
    doi = "10.1103/PhysRevLett.122.221301",
    journal = "Phys. Rev. Lett.",
    volume = "122",
    number = "22",
    pages = "221301",
    year = "2019"
}

@article{Smith:2019ihp,
    author = "Smith, Tristan L. and Poulin, Vivian and Amin, Mustafa A.",
    title = "{Oscillating scalar fields and the Hubble tension: a resolution with novel signatures}",
    eprint = "1908.06995",
    archivePrefix = "arXiv",
    primaryClass = "astro-ph.CO",
    doi = "10.1103/PhysRevD.101.063523",
    journal = "Phys. Rev. D",
    volume = "101",
    number = "6",
    pages = "063523",
    year = "2020"
}

@article{Agrawal:2019lmo,
    author = "Agrawal, Prateek and Cyr-Racine, Francis-Yan and Pinner, David and Randall, Lisa",
    title = "{Rock {\textquoteleft}n{\textquoteright} roll solutions to the Hubble tension}",
    eprint = "1904.01016",
    archivePrefix = "arXiv",
    primaryClass = "astro-ph.CO",
    doi = "10.1016/j.dark.2023.101347",
    journal = "Phys. Dark Univ.",
    volume = "42",
    pages = "101347",
    year = "2023"
}

@article{Lin:2019qug,
    author = "Lin, Meng-Xiang and Benevento, Giampaolo and Hu, Wayne and Raveri, Marco",
    title = "{Acoustic Dark Energy: Potential Conversion of the Hubble Tension}",
    eprint = "1905.12618",
    archivePrefix = "arXiv",
    primaryClass = "astro-ph.CO",
    doi = "10.1103/PhysRevD.100.063542",
    journal = "Phys. Rev. D",
    volume = "100",
    number = "6",
    pages = "063542",
    year = "2019"
}

@article{Niedermann:2019olb,
    author = "Niedermann, Florian and Sloth, Martin S.",
    title = "{New early dark energy}",
    eprint = "1910.10739",
    archivePrefix = "arXiv",
    primaryClass = "astro-ph.CO",
    doi = "10.1103/PhysRevD.103.L041303",
    journal = "Phys. Rev. D",
    volume = "103",
    number = "4",
    pages = "L041303",
    year = "2021"
}

@article{Niedermann:2020dwg,
    author = "Niedermann, Florian and Sloth, Martin S.",
    title = "{Resolving the Hubble tension with new early dark energy}",
    eprint = "2006.06686",
    archivePrefix = "arXiv",
    primaryClass = "astro-ph.CO",
    doi = "10.1103/PhysRevD.102.063527",
    journal = "Phys. Rev. D",
    volume = "102",
    number = "6",
    pages = "063527",
    year = "2020"
}

@article{Karwal:2021vpk,
    author = "Karwal, Tanvi and Raveri, Marco and Jain, Bhuvnesh and Khoury, Justin and Trodden, Mark",
    title = "{Chameleon early dark energy and the Hubble tension}",
    eprint = "2106.13290",
    archivePrefix = "arXiv",
    primaryClass = "astro-ph.CO",
    doi = "10.1103/PhysRevD.105.063535",
    journal = "Phys. Rev. D",
    volume = "105",
    number = "6",
    pages = "063535",
    year = "2022"
}

@article{McDonough:2021pdg,
    author = "McDonough, Evan and Lin, Meng-Xiang and Hill, J. Colin and Hu, Wayne and Zhou, Shengjia",
    title = "{Early dark sector, the Hubble tension, and the swampland}",
    eprint = "2112.09128",
    archivePrefix = "arXiv",
    primaryClass = "astro-ph.CO",
    doi = "10.1103/PhysRevD.106.043525",
    journal = "Phys. Rev. D",
    volume = "106",
    number = "4",
    pages = "043525",
    year = "2022"
}

@article{Lin:2022phm,
    author = "Lin, Meng-Xiang and McDonough, Evan and Hill, J. Colin and Hu, Wayne",
    title = "{Dark matter trigger for early dark energy coincidence}",
    eprint = "2212.08098",
    archivePrefix = "arXiv",
    primaryClass = "astro-ph.CO",
    doi = "10.1103/PhysRevD.107.103523",
    journal = "Phys. Rev. D",
    volume = "107",
    number = "10",
    pages = "103523",
    year = "2023"
}

@article{Vagnozzi:2021gjh,
    author = "Vagnozzi, Sunny",
    title = "{Consistency tests of {\ensuremath{\Lambda}}CDM from the early integrated Sachs-Wolfe effect: Implications for early-time new physics and the Hubble tension}",
    eprint = "2105.10425",
    archivePrefix = "arXiv",
    primaryClass = "astro-ph.CO",
    doi = "10.1103/PhysRevD.104.063524",
    journal = "Phys. Rev. D",
    volume = "104",
    number = "6",
    pages = "063524",
    year = "2021"
}

@article{Hill:2020osr,
    author = "Hill, J. Colin and McDonough, Evan and Toomey, Michael W. and Alexander, Stephon",
    title = "{Early dark energy does not restore cosmological concordance}",
    eprint = "2003.07355",
    archivePrefix = "arXiv",
    primaryClass = "astro-ph.CO",
    doi = "10.1103/PhysRevD.102.043507",
    journal = "Phys. Rev. D",
    volume = "102",
    number = "4",
    pages = "043507",
    year = "2020"
}

@article{Ivanov:2020ril,
    author = "Ivanov, Mikhail M. and McDonough, Evan and Hill, J. Colin and Simonovi{\'c}, Marko and Toomey, Michael W. and Alexander, Stephon and Zaldarriaga, Matias",
    title = "{Constraining Early Dark Energy with Large-Scale Structure}",
    eprint = "2006.11235",
    archivePrefix = "arXiv",
    primaryClass = "astro-ph.CO",
    reportNumber = "INR-TH-2020-035; CERN-TH-2020-103",
    doi = "10.1103/PhysRevD.102.103502",
    journal = "Phys. Rev. D",
    volume = "102",
    number = "10",
    pages = "103502",
    year = "2020"
}

@article{DAmico:2020ods,
    author = "D'Amico, Guido and Senatore, Leonardo and Zhang, Pierre and Zheng, Henry",
    title = "{The Hubble Tension in Light of the Full-Shape Analysis of Large-Scale Structure Data}",
    eprint = "2006.12420",
    archivePrefix = "arXiv",
    primaryClass = "astro-ph.CO",
    doi = "10.1088/1475-7516/2021/05/072",
    journal = "JCAP",
    volume = "05",
    pages = "072",
    year = "2021"
}

@article{Murgia:2020ryi,
    author = "Murgia, Riccardo and Abell{\'a}n, Guillermo F. and Poulin, Vivian",
    title = "{Early dark energy resolution to the Hubble tension in light of weak lensing surveys and lensing anomalies}",
    eprint = "2009.10733",
    archivePrefix = "arXiv",
    primaryClass = "astro-ph.CO",
    doi = "10.1103/PhysRevD.103.063502",
    journal = "Phys. Rev. D",
    volume = "103",
    number = "6",
    pages = "063502",
    year = "2021"
}

@article{Goldstein:2023gnw,
    author = "Goldstein, Samuel and Hill, J. Colin and Ir{\v{s}}i{\v{c}}, Vid and Sherwin, Blake D.",
    title = "{Canonical Hubble-Tension-Resolving Early Dark Energy Cosmologies Are Inconsistent with the Lyman-{\ensuremath{\alpha}} Forest}",
    eprint = "2303.00746",
    archivePrefix = "arXiv",
    primaryClass = "astro-ph.CO",
    doi = "10.1103/PhysRevLett.131.201001",
    journal = "Phys. Rev. Lett.",
    volume = "131",
    number = "20",
    pages = "201001",
    year = "2023"
}

@article{Gsponer:2023wpm,
    author = "Gsponer, Rafaela and Zhao, Ruiyang and Donald-McCann, Jamie and Bacon, David and Koyama, Kazuya and Crittenden, Robert and Simon, Theo and Mueller, Eva-Maria",
    title = "{Cosmological constraints on early dark energy from the full shape analysis of eBOSS DR16}",
    eprint = "2312.01977",
    archivePrefix = "arXiv",
    primaryClass = "astro-ph.CO",
    doi = "10.1093/mnras/stae992",
    journal = "Mon. Not. Roy. Astron. Soc.",
    volume = "530",
    number = "3",
    pages = "3075--3099",
    year = "2024"
}

@article{Herold:2021ksg,
    author = "Herold, Laura and Ferreira, Elisa G. M. and Komatsu, Eiichiro",
    title = "{New Constraint on Early Dark Energy from Planck and BOSS Data Using the Profile Likelihood}",
    eprint = "2112.12140",
    archivePrefix = "arXiv",
    primaryClass = "astro-ph.CO",
    doi = "10.3847/2041-8213/ac63a3",
    journal = "Astrophys. J. Lett.",
    volume = "929",
    number = "1",
    pages = "L16",
    year = "2022"
}

@article{Herold:2022iib,
    author = "Herold, Laura and Ferreira, Elisa G. M.",
    title = "{Resolving the Hubble tension with early dark energy}",
    eprint = "2210.16296",
    archivePrefix = "arXiv",
    primaryClass = "astro-ph.CO",
    doi = "10.1103/PhysRevD.108.043513",
    journal = "Phys. Rev. D",
    volume = "108",
    number = "4",
    pages = "043513",
    year = "2023"
}

@article{Smith:2023oop,
    author = "Smith, Tristan L. and Poulin, Vivian",
    title = "{Current small-scale CMB constraints to axionlike early dark energy}",
    eprint = "2309.03265",
    archivePrefix = "arXiv",
    primaryClass = "astro-ph.CO",
    doi = "10.1103/PhysRevD.109.103506",
    journal = "Phys. Rev. D",
    volume = "109",
    number = "10",
    pages = "103506",
    year = "2024"
}

@article{Efstathiou:2023fbn,
    author = "Efstathiou, George and Rosenberg, Erik and Poulin, Vivian",
    title = "{Improved Planck Constraints on Axionlike Early Dark Energy as a Resolution of the Hubble Tension}",
    eprint = "2311.00524",
    archivePrefix = "arXiv",
    primaryClass = "astro-ph.CO",
    doi = "10.1103/PhysRevLett.132.221002",
    journal = "Phys. Rev. Lett.",
    volume = "132",
    number = "22",
    pages = "221002",
    year = "2024"
}

@article{Khalife:2023qbu,
    author = {Khalife, Ali Rida and Zanjani, Maryam Bahrami and Galli, Silvia and G{\"u}nther, Sven and Lesgourgues, Julien and Benabed, Karim},
    title = "{Review of Hubble tension solutions with new SH0ES and SPT-3G data}",
    eprint = "2312.09814",
    archivePrefix = "arXiv",
    primaryClass = "astro-ph.CO",
    reportNumber = "TTK-23-36",
    doi = "10.1088/1475-7516/2024/04/059",
    journal = "JCAP",
    volume = "04",
    pages = "059",
    year = "2024"
}

@article{Giare:2024akf,
    author = "Giar{\`e}, William",
    title = "{Inflation, the Hubble tension, and early dark energy: An alternative overview}",
    eprint = "2404.12779",
    archivePrefix = "arXiv",
    primaryClass = "astro-ph.CO",
    doi = "10.1103/PhysRevD.109.123545",
    journal = "Phys. Rev. D",
    volume = "109",
    number = "12",
    pages = "123545",
    year = "2024"
}

@article{Strunk:2026vve,
    author = "Strunk, J. and others",
    title = "{The SRG/eROSITA All-Sky Survey: Early dark energy and Hubble constant from the cluster mass function}",
    eprint = "2609.07834",
    archivePrefix = "arXiv",
    primaryClass = "astro-ph.CO",
    month = "9",
    year = "2026"
}

@article{Jhaveri:2026bla,
    author = "Jhaveri, Tanisha and Karwal, Tanvi and Crawford, Thomas and Hu, Wayne and Khalife, Ali Rida and Balkenhol, Lennart and Ge, Fei",
    title = "{Disentangling cosmic distance tensions with early and late dark energy}",
    eprint = "2604.08530",
    archivePrefix = "arXiv",
    primaryClass = "astro-ph.CO",
    month = "4",
    year = "2026"
}

@article{Caldwell:2005tm,
    author = "Caldwell, R. R. and Linder, Eric V.",
    title = "{The Limits of quintessence}",
    eprint = "astro-ph/0505494",
    archivePrefix = "arXiv",
    doi = "10.1103/PhysRevLett.95.141301",
    journal = "Phys. Rev. Lett.",
    volume = "95",
    pages = "141301",
    year = "2005"
}

@article{Braglia:2020bym,
    author = "Braglia, Matteo and Emond, William T. and Finelli, Fabio and Gumrukcuoglu, A. Emir and Koyama, Kazuya",
    title = "{Unified framework for early dark energy from $\alpha$-attractors}",
    eprint = "2005.14053",
    archivePrefix = "arXiv",
    primaryClass = "astro-ph.CO",
    doi = "10.1103/PhysRevD.102.083513",
    journal = "Phys. Rev. D",
    volume = "102",
    number = "8",
    pages = "083513",
    year = "2020"
}

@article{Brissenden:2023yko,
    author = "Brissenden, Lucy and Dimopoulos, Konstantinos and S{\'a}nchez L{\'o}pez, Samuel",
    title = "{Non-oscillating early dark energy and quintessence from {\ensuremath{\alpha}}-attractors}",
    eprint = "2301.03572",
    archivePrefix = "arXiv",
    primaryClass = "astro-ph.CO",
    doi = "10.1016/j.astropartphys.2024.102925",
    journal = "Astropart. Phys.",
    volume = "157",
    pages = "102925",
    year = "2024"
}

@article{Freese:2021rjq,
    author = "Freese, Katherine and Winkler, Martin Wolfgang",
    title = "{Chain early dark energy: A Proposal for solving the Hubble tension and explaining today{\textquoteright}s dark energy}",
    eprint = "2102.13655",
    archivePrefix = "arXiv",
    primaryClass = "astro-ph.CO",
    reportNumber = "UTTG-02-2021, NORDITA-2021-020",
    doi = "10.1103/PhysRevD.104.083533",
    journal = "Phys. Rev. D",
    volume = "104",
    number = "8",
    pages = "083533",
    year = "2021"
}

@article{Ramadan:2023ivw,
    author = "Ramadan, Omar F. and Karwal, Tanvi and Sakstein, Jeremy",
    title = "{Attractive proposal for resolving the Hubble tension: Dynamical attractors that unify early and late dark energy}",
    eprint = "2309.08082",
    archivePrefix = "arXiv",
    primaryClass = "astro-ph.CO",
    doi = "10.1103/PhysRevD.109.063525",
    journal = "Phys. Rev. D",
    volume = "109",
    number = "6",
    pages = "063525",
    year = "2024"
}

@article{Guendelman:2022cop,
    author = "Guendelman, Eduardo and Herrera, Ram{\'o}n and Benisty, David",
    title = "{Unifying inflation with early and late dark energy with multiple fields: Spontaneously broken scale invariant two measures theory}",
    eprint = "2201.06470",
    archivePrefix = "arXiv",
    primaryClass = "gr-qc",
    doi = "10.1103/PhysRevD.105.124035",
    journal = "Phys. Rev. D",
    volume = "105",
    number = "12",
    pages = "124035",
    year = "2022"
}

@article{Nojiri:2023mvi,
    author = "Nojiri, Shin'ichi and Odintsov, Sergei D. and S{\'a}ez-Chill{\'o}n G{\'o}mez, Diego",
    title = "{Unifying inflation with early and late dark energy in Einstein{\textendash}Gauss{\textendash}Bonnet gravity}",
    eprint = "2304.08255",
    archivePrefix = "arXiv",
    primaryClass = "gr-qc",
    doi = "10.1016/j.dark.2023.101238",
    journal = "Phys. Dark Univ.",
    volume = "41",
    pages = "101238",
    year = "2023"
}

@article{Giare:2026tyk,
    author = "Giar{\`e}, William and Sakstein, Jeremy",
    title = "{Unifying Early and Late Dark Energy: Dynamical Requirements and Obstructions}",
    eprint = "2605.26116",
    archivePrefix = "arXiv",
    primaryClass = "astro-ph.CO",
    month = "5",
    year = "2026"
}

@article{Fields:2019pfx,
    author = "Fields, Brian D. and Olive, Keith A. and Yeh, Tsung-Han and Young, Charles",
    title = "{Big-Bang Nucleosynthesis after Planck}",
    eprint = "1912.01132",
    archivePrefix = "arXiv",
    primaryClass = "astro-ph.CO",
    reportNumber = "UMN--TH--3902/19, FTPI--MINN--19/25",
    doi = "10.1088/1475-7516/2020/03/010",
    journal = "JCAP",
    volume = "03",
    pages = "010",
    year = "2020",
    note = "[Erratum: JCAP 11, E02 (2020)]"
}

@article{Abdalla:2022yfr,
    author = "Abdalla, Elcio and others",
    title = "{Cosmology intertwined: A review of the particle physics, astrophysics, and cosmology associated with the cosmological tensions and anomalies}",
    eprint = "2203.06142",
    archivePrefix = "arXiv",
    primaryClass = "astro-ph.CO",
    reportNumber = "FERMILAB-CONF-22-192-SCD",
    doi = "10.1016/j.jheap.2022.04.002",
    journal = "JHEAp",
    volume = "34",
    pages = "49--211",
    year = "2022"
}

@article{DES:2026fyc,
    author = "Abbott, T. M. C. and others",
    collaboration = "DES",
    title = "{Dark Energy Survey Year 6 Results: Cosmological Constraints from Galaxy Clustering and Weak Lensing}",
    eprint = "2601.14559",
    archivePrefix = "arXiv",
    primaryClass = "astro-ph.CO",
    reportNumber = "DES-2025-0929, FERMILAB-PUB-26-0026-PPD",
    month = "1",
    year = "2026"
}

@article{Planck:2019nip,
    author = "Aghanim, N. and others",
    collaboration = "Planck",
    title = "{Planck 2018 results. V. CMB power spectra and likelihoods}",
    eprint = "1907.12875",
    archivePrefix = "arXiv",
    primaryClass = "astro-ph.CO",
    doi = "10.1051/0004-6361/201936386",
    journal = "Astron. Astrophys.",
    volume = "641",
    pages = "A5",
    year = "2020"
}

@article{Ma:1995ey,
    author = "Ma, Chung-Pei and Bertschinger, Edmund",
    title = "{Cosmological perturbation theory in the synchronous
              and conformal Newtonian gauges}",
    eprint = "astro-ph/9506072",
    archivePrefix = "arXiv",
    doi = "10.1086/176550",
    journal = "Astrophys. J.",
    volume = "455",
    pages = "7--25",
    year = "1995"
}

@article{Lesgourgues:2011re,
    author = "Lesgourgues, Julien",
    title = "{The Cosmic Linear Anisotropy Solving System
              (CLASS) I: Overview}",
    eprint = "1104.2932",
    archivePrefix = "arXiv",
    primaryClass = "astro-ph.IM",
    reportNumber = "LAPTH-009-11",
    year = "2011"
}

@article{Blas:2011rf,
    author = "Blas, Diego and Lesgourgues, Julien and Tram, Thomas",
    title = "{The Cosmic Linear Anisotropy Solving System
              (CLASS). Part II: Approximation schemes}",
    eprint = "1104.2933",
    archivePrefix = "arXiv",
    primaryClass = "astro-ph.CO",
    doi = "10.1088/1475-7516/2011/07/034",
    journal = "JCAP",
    volume = "07",
    pages = "034",
    year = "2011"
}

@article{AtacamaCosmologyTelescope:2025blo,
    author = "Louis, Thibaut and others",
    collaboration = "Atacama Cosmology Telescope",
    title = "{The Atacama Cosmology Telescope: DR6 power spectra,
              likelihoods and {\ensuremath{\Lambda}}CDM parameters}",
    eprint = "2503.14452",
    archivePrefix = "arXiv",
    primaryClass = "astro-ph.CO",
    doi = "10.1088/1475-7516/2025/11/062",
    journal = "JCAP",
    volume = "11",
    pages = "062",
    year = "2025"
}

@article{SPT-3G:2025bzu,
    author = "Camphuis, E. and others",
    collaboration = "SPT-3G",
    title = "{SPT-3G D1: CMB temperature and polarization power spectra
              and cosmology from 2019 and 2020 observations
              of the SPT-3G main field}",
    eprint = "2506.20707",
    archivePrefix = "arXiv",
    primaryClass = "astro-ph.CO",
    doi = "10.1103/7wt3-9v2y",
    journal = "Phys. Rev. D",
    volume = "113",
    number = "8",
    pages = "083504",
    year = "2026"
}

@article{SPT-3G:2024atg,
    author = "Ge, F. and others",
    collaboration = "SPT-3G",
    title = "{Cosmology from CMB lensing and delensed EE power spectra
              using 2019{\textendash}2020 SPT-3G polarization data}",
    eprint = "2411.06000",
    archivePrefix = "arXiv",
    primaryClass = "astro-ph.CO",
    doi = "10.1103/PhysRevD.111.083534",
    journal = "Phys. Rev. D",
    volume = "111",
    number = "8",
    pages = "083534",
    year = "2025"
}

@article{Balkenhol:2024sbv,
    author = "Balkenhol, L. and Trendafilova, C.
              and Benabed, K. and Galli, S.",
    title = "{candl: cosmic microwave background analysis
              with a differentiable likelihood}",
    eprint = "2401.13433",
    archivePrefix = "arXiv",
    primaryClass = "astro-ph.CO",
    doi = "10.1051/0004-6361/202449432",
    journal = "Astron. Astrophys.",
    volume = "686",
    pages = "A10",
    year = "2024"
}

@article{Torrado:2020dgo,
    author = "Torrado, Jesus and Lewis, Antony",
    title = "{Cobaya: Code for Bayesian Analysis of hierarchical physical models}",
    eprint = "2005.05290",
    archivePrefix = "arXiv",
    primaryClass = "astro-ph.IM",
    doi = "10.1088/1475-7516/2021/05/057",
    journal = "JCAP",
    volume = "05",
    pages = "057",
    year = "2021"
}

@article{Mead:2020vgs,
    author = {Mead, Alexander and Brieden, Samuel
              and Tr{\"o}ster, Tilman and Heymans, Catherine},
    title = "{hmcode-2020: improved modelling of non-linear
              cosmological power spectra with baryonic feedback}",
    eprint = "2009.01858",
    archivePrefix = "arXiv",
    primaryClass = "astro-ph.CO",
    doi = "10.1093/mnras/stab082",
    journal = "Mon. Not. Roy. Astron. Soc.",
    volume = "502",
    number = "1",
    pages = "1401--1422",
    year = "2021"
}

@article{Sohail:2024oki,
    author = "Sohail, Sk. and Alam, Sonej and Akthar, Shiriny and Hossain, Md. Wali",
    title = "{Quintessential early dark energy}",
    eprint = "2408.03229",
    archivePrefix = "arXiv",
    primaryClass = "astro-ph.CO",
    doi = "10.1016/j.dark.2025.101948",
    journal = "Phys. Dark Univ.",
    volume = "48",
    pages = "101948",
    year = "2025"
}

@article{Kouniatalis:2026upo,
    author = "Kouniatalis, Gerasimos",
    title = "{A Unified Scalar-Field Mechanism for the $H_0$, $S_8$ and Evolving Dark-Energy Tensions}",
    eprint = "2603.11706",
    archivePrefix = "arXiv",
    primaryClass = "astro-ph.CO",
    doi = "10.3390/astronomy5040017",
    journal = "Astronomy",
    volume = "5",
    pages = "17",
    year = "2026"
}

@article{Guendelman:2025swp,
    author = "Guendelman, Eduardo I. and Herrera, Ramon and Labrana, Pedro",
    title = "{Connecting early dark energy to late dark energy by the diluting matter potential}",
    eprint = "2507.17095",
    archivePrefix = "arXiv",
    primaryClass = "gr-qc",
    doi = "10.1140/epjc/s10052-026-15794-x",
    journal = "Eur. Phys. J. C",
    volume = "86",
    number = "5",
    pages = "571",
    year = "2026"
}

@article{Dong:2026epd,
    author = "Dong, Haotian and Biesiada, Marek and Cao, Zhoujian and Wang, Shao-Jiang",
    title = "{A cosmological intercept tension in local distance ladders}",
    eprint = "2610.31181",
    archivePrefix = "arXiv",
    primaryClass = "astro-ph.CO",
    month = "10",
    year = "2026"
}

@article{Ling:2026awm,
    author = "Ling, Jia-Le and Wang, Shao-Jiang",
    title = "{GPAge: a Global Parameterization based on cosmic Age for some unified early-time and late-time solutions to the Hubble tension}",
    eprint = "2610.30961",
    archivePrefix = "arXiv",
    primaryClass = "astro-ph.CO",
    month = "10",
    year = "2026"
}

@article{Giare:2026gdn,
    author = "Giar{\`e}, William and Sakstein, Jeremy",
    title = "{Early Dark Energy Crowds Out Interacting Dark Energy}",
    eprint = "2610.37836",
    archivePrefix = "arXiv",
    primaryClass = "astro-ph.CO",
    month = "10",
    year = "2026"
}

@article{Jones:2026ndt,
    author = "Jones, Joann and Uch{\^o}a, Ualype de Andrade and Dodelson, Scott",
    title = "{Disentangling constraints on $S_8$ from the high and low redshift CMB}",
    eprint = "2610.37752",
    archivePrefix = "arXiv",
    primaryClass = "astro-ph.CO",
    month = "10",
    year = "2026"
}

@article{Najera:2026odk,
    author = "N{\'a}jera, Jos{\'e} Antonio and Banik, Indranil and Desmond, Harry",
    title = "{Does the inverse distance ladder rule out late-time solutions to the Hubble tension? A model-independent test using DES-Dovekie supernovae and DESI DR2 BAO}",
    eprint = "2610.23657",
    archivePrefix = "arXiv",
    primaryClass = "astro-ph.CO",
    month = "10",
    year = "2026"
}

\end{document}